\documentclass[aps,pre,reprint,floatfix]{revtex4-1}

\usepackage{graphicx}
\graphicspath{{Images_paper/}}
\DeclareGraphicsExtensions{.pdf,.png}

\usepackage{grffile}

\usepackage{xcolor}
\usepackage{amsmath}
\usepackage{amsthm, amssymb, amsfonts, amsbsy}
\usepackage{mathtools} 
\usepackage{bm}
\usepackage{mathrsfs}  
\usepackage{tipa}

\usepackage{tikz}
\usepackage{tikz-3dplot}
\usepackage{subcaption}

\usepackage[T1]{fontenc}
\usepackage[utf8]{inputenc}

\usepackage{textcomp}

\usepackage[english]{babel}

\newcommand{\ds}{\displaystyle}
\newcommand{\beq}{\begin{equation}}
\newcommand{\eeq}{\end{equation}}
\newcommand{\beqa}{\begin{eqnarray}}
\newcommand{\eeqa}{\end{eqnarray}}
\newcommand{\bem}{\begin{math}}
\newcommand{\eem}{\end{math}}
\newcommand{\rar}{{\rightarrow}}

\newcommand{\overbar}[1]{\mkern 1.5mu\overline{\mkern-1.5mu#1\mkern-1.5mu}\mkern 1.5mu}

\newcommand{\ham}{{\cal H}}
\newcommand{\ent}{{\cal S}}
\newcommand{\bmp}{{\bm p}}
\newcommand{\bmP}{{\bm P}}
\newcommand{\bmq}{{\bm q}}
\newcommand{\bmQ}{{\bm Q}}
\newcommand{\bmn}{{\bm n}}
\newcommand{\bmA}{{\bm A}}
\newcommand{\bmS}{{\bm S}}
\newcommand{\bms}{{\bm s}}
\newcommand{\bmI}{{\bm I}}

\newcommand{\bfr}{{\bf r}}
\newcommand{\bft}{{\bf t}}

\newcommand{\bfn}{{\bf n}}

\newcommand{\bfR}{{\bf R}}

\newcommand{\aver}[1]{\left\langle {#1}\right\rangle}
\newcommand{\mean}[1]{ \overbar{{#1}} }

\newcommand{\ddt}[1]{{\ds { d {#1} \over d t}}}

\newcommand{\norm}[1]{\left \lVert #1 \right \rVert}

\usepackage[capitalise]{cleveref}
\crefformat{equation}{Eq. (#2#1#3)} 
\Crefformat{equation}{Equation (#2#1#3)} 
\crefrangeformat{equation}{Eqs. (#3#1#4--#5\crefstripprefix{#1}{#2}#6)} 
\Crefrangeformat{equation}{Equations (#3#1#4--#5\crefstripprefix{#1}{#2}#6)} 
\crefmultiformat{equation}{Eqs. (#2#1#3)}{ and~(#2#1#3)}{, (#2#1#3)}{ and~(#2#1#3)} 
\Crefmultiformat{equation}{Equations (#2#1#3)}{ and~(#2#1#3)}{, (#2#1#3)}{ and~(#2#1#3)} 

\usepackage{xcolor}

\def\tl#1{\textcolor{black}{#1}}

\begin{document}


\title{
Fluctuations of cell geometry and their non-equilibrium thermodynamics in living epithelial tissue 
} 

\author{M. Olenik $^{1,*}$}
\author{J. Turley $^{1,2,*}$}
\author{S. Cross $^2$}
\author{H. Weavers $^2$}
\author{P. Martin $^2$}
\author{I.V. Chenchiah $^{1}$}
\author{T. B. Liverpool $^{1}$}

\affiliation{School of Mathematics, University of Bristol - Bristol BS8 1UG, UK}

\affiliation{School of Biochemistry, University of Bristol - Bristol BS8 1TW, UK}

\date{\today}

\begin{abstract} 
We \tl{measure different contributions to entropy production in a living functional epithelial tissue. We do this by extracting the functional dynamics of development while at the same time quantifying  fluctuations. Using the translucent {\em Drosophila melanogaster} pupal epithelium as an ideal tissue for high resolution live imaging~\cite{weavers2018long},  we measure the entropy associated with the stochastic geometry of cells in the epithelium. This is done using a detailed analysis of the dynamics of the shape and orientation of individual cells which enables separation of local and global aspects of the tissue behaviour. We find intriguingly that we can observe irreversible dynamics in the cell geometries but without a change in the entropy associated with those degrees of freedom, showing that there is a flow of energy into those degrees of freedom. Hence the living system is controlling how the entropy is being produced and partitioned into its different parts.}  
\end{abstract}

\maketitle


\section{Introduction}

A tissue is a group of similar cells that function together as a unit. Hence there is a hope that ideas and techniques to describe many particle systems from condensed matter physics will be helpful to understand their function~\cite{Andersen1972}. There have already been some significant successes following this line of reasoning.
The mechanical influences on the dynamic interplay between cells in epithelial tissues have been shown to be important for a diverse array of biological processes from embryonic development and growth~\cite{Lecuit2007a,Lecuit2011} through to healing of wounds~\cite{Martin1992,Razzell2014a,Tetley2019,turley2022good} and other pathologies  like cancer \cite{piersma2020fibrosis}.  Due to their  importance in understanding tissues there has been much work on quantifying and inferring these forces \cite{hutson2003forces, ishihara2012bayesian, sugimura2013mechanical, etournay2015interplay}.
%

However self-sustaining tissue is different from a collection of cells (its constituent parts) due to a variety of stochastic~\cite{Elowitz2002,Chubb2010} feedback processes and information flow essential for life to function. More is different but in not quite the same way as in traditional  condensed matter~\cite{Andersen1972}.
To consider the physical constraints on biological function in the most realistic context, our focus here is quantifying
{\em functional} living tissue~\cite{Aigouy2010a} {\em in-vivo}. 
By doing this we aim to contrast with recent work on collective cell behaviour in
{\em in-vitro} sheets  of cells~\cite{Duclos2014,Duclos2017b,Li2017,Saw2017,Kawaguchi2017,Martella2019} or {\em ex-vivo} tissue extracts~\cite{Tetley2019}. 
We explore this difference  by quantifying the fluctuations as well as the dynamics of various geometric quantities in living tissue at cellular scales looking for signatures of functionality~\cite{Brenner2012}.
%
%
From analysis of the fluctuations 
we measure the system's entropy and how it evolves in time.
The fact that entropy increases for irreversible adiabatic processes (the 2nd law of thermodynamics) is one of the touchstones of modern physics.  \tl{It is not clear if entropy is even a useful concept in living multicellular organisms which of course are {\em not} adiabatic, nor at equilibrium. In addition, 
the question of how entropy evolves 
in living multicellular organisms  
remains an open question which we try to address here. A key part of our analysis involves defining a {\em partial} entropy~\cite{Ahmadzade2017}, depending only on a subset of the total number of degrees of freedom (d.o.f) available. The other degrees of freedom can then be viewed as a part of a reservoir in contact with these d.o.f.'s.} 

We \tl{also
precisely  quantify} a well-known analogy~\cite{Bouligand2008} to the breaking of rotational symmetry occurring in the transitions from an isotropic to an ordered mesophase in a liquid crystal~\cite{DeGennesProst93} and the global shape and orientational order occurring in regions of developing tissue~\cite{Duvert1984}. We map the shape and orientations of  cells in the epithelium to a model liquid crystal. While easy to see for in-vitro  sheets of confluent elongated cells like fibroblasts~\cite{Duclos2014,Duclos2017b}, it is harder to make this analogy for the more isotropic cells found in functional tissue. This also addresses the  question whether the cells show nematic~\cite{Olmsted1990} or polar (ferroelectric) ~\cite{Chen2020}  liquid crystalline order. We show unambiguously that epithelial cells \tl{in a living functional tissue} show nematic order on large scales but polar order on very small scales which quickly decays to zero over a correlation length comparable to the size of a single cell. Furthermore, the irreversible process~\cite{Onsager1931b,Fang2019} of epithelial growth is indicated by the amount of liquid crystal (nematic) order increasing with time while at the same time, the associated observed {\em partial} (information) entropy~\cite{Plischke2005,Bialek2011} remains constant. We study epithelial tissue at a well characterised stage of development where the the hinge region is contracting stretching the wing blade. We are investigating this focusing on the fluctuations on the cellular scale \cite{athilingam2021mechanics, aigouy2010cell}. We show that throughout a one hour period of observation, despite the increasing average orientational order that the probability distribution of cell shapes and orientations follows a universal form which remains steady, but that while the individual cells behaviour fluctuates widely, their average behaviour 
evolves in a precise deterministic way according to an equation of motion which we are able to obtain. This implies that the average dynamics are tightly coupled to the fluctuations and vice versa providing a special type of steady state where the density of states remains constant even while the system is evolving continually in time~\cite{Liverpool2020}.
This suggests that zero entropy production, in the presence of irreversible dynamics,  
can be a way to identify healthy functional living tissue. 
Using this entropy, we are
able to separate the fluctuations~\cite{Elowitz2002,Raser2005} in the tissue at the cellular level from the large scale changes occurring that lead to development, growth and morphogenesis. 

\begin{figure*}[h!] 
\begin{center}

	\includegraphics[scale=0.65]{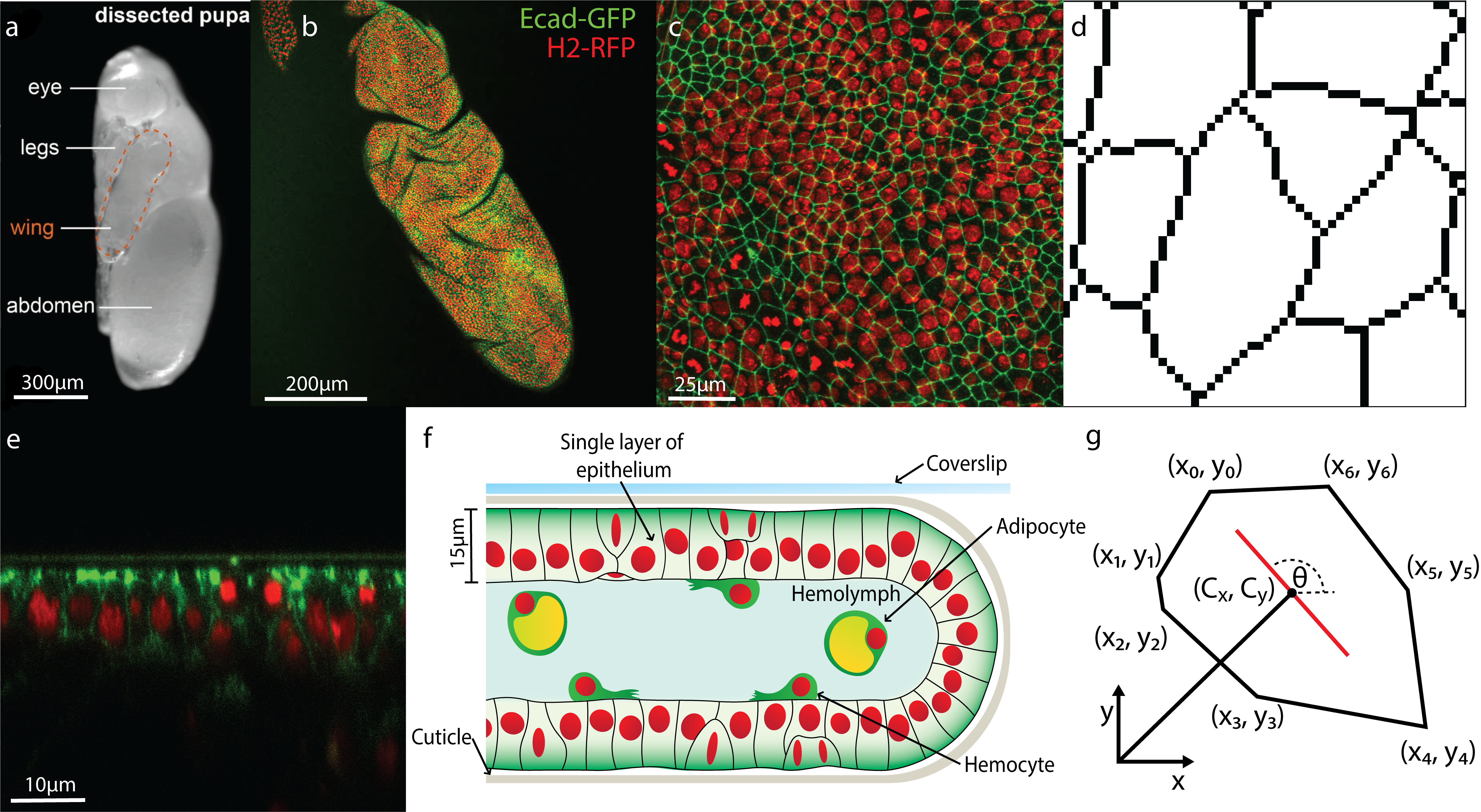} 
	\caption{\small \sl a) Translucent Drosophila pupa after dissection from its puparium~\cite{weavers2018long,thuma2018drosophila}. b) Low magnification of the entire transgenic pupal wing; the total wing length is 800$\mu$m and the width is 250$\mu$m. The green is "Ecadherin-GFP" which labels the cell's "Adherens Junctions" and the red is "Histone-RFP" which labels the cell nuclei. c) Pupal wing tissue. d) A high magnification (binary) view of cell boundaries. e)  High magnification cross-section of the wing.  f) A schematic diagram of the cross-section of the wing. g) Geometric characterisation of a cell and its boundary. Scale bars as indicated.
	} \label{fig:full_wing}
\end{center}
\end{figure*}

\textit{Drosophila} is highly genetically tractable and is well characterised as a model of embryonic development and disease~\cite{razzell2011swatting}.  Here we study the pupal stage of its life cycle because pupae are translucent and immobile, (see fig. \ref{fig:full_wing}a).  
More specifically, we investigate pupal wings, (see fig. \ref{fig:full_wing}b and fig. \ref{fig:full_wing}f), which enable viewing of  a flat 2D surface of epithelial cells, to gather data rich high resolution in-vivo images with relative ease~\cite{weavers2018long}. These images of the wing provide us with a unique opportunity for a level of analysis  that is currently not possible in other systems due to their opacity and the difficulty in obtaining equivalent quality {\em in-vivo} data. 

Confocal movies were captured from 
18 hours post pupal formation.  The pupae are first dissected and removed from their puparium. This allowed us to directly view the wing, (see fig. \ref{fig:full_wing}f), and to gather confocal time-lapse movies of the living wing tissue, details of which are in the supplementary. An example image from a movie is given in fig \ref{fig:full_wing}c, with a cross-sectional view in fig. \ref{fig:full_wing}e. From the experimental data we generate binary images (as shown in fig. \ref{fig:full_wing}d \cite{rueden2017imagej2, schindelin2012fiji, stephen_cross_2018_2525263,Umorin_2002}. 
The cell boundaries are fitted to polygons ($\sim$ 5-16 edges   and where a single cell-cell boundary can have more than one edge) suitable for efficient  mathematical analysis (see fig. \ref{fig:full_wing}g).

\onecolumngrid

\twocolumngrid

\tl{Using these movies we can perform a detailed statistical analysis of the evolution in time of the shapes, sizes and orientations of cells looking not only on their typical behaviour but also their fluctuations. This will allow us to maximise the amount of useful information that we can extract  from these observations. Hence this will put is in the position to most accurately quantify what happens to such tissues after perturbations such as wounds.
}


\section{Methods}

\paragraph{Experimental Methods}
Transgenic \textit{Drosophila} were used that expressed the endogenous Ecadherin protein artificially tagged with GFP (green fluorescent protein).
E-cadherin is a cell-cell adhesion molecule, this means the fluorescent protein is localised at the cell membranes allowing easy identification of cell-cell boundaries via microscopy (see fig. \ref{fig:full_wing}c). Images were taken of pupae at 18 hours after puparium formation (APF). To gain images of the pupae wings, first the pupae were dissected to remove the puparium case. This allows us to directly view the wing \cite{weavers2018long}. The pupae are transferred to an imaging dish with the wing in direct contact with the coverglass. Wings from transgenic \textit{Drosophila} pupae labelled with Ecadherin-GFP and Histone-RFP are imaged on a Leica SP8 confocal microscope. Z-stacks are taken through the 3D wing (with z-slices at 0.75 $\mu$m intervals) and images are taken every 5mins.  This produces a time-lapse video of the healthy, developing tissue. Each image consists of a series of different heights through the tissue ($Z$-stacks) \cite{weavers2018long}. 
The individual $3D$ stacks are transformed into a 2D images using a modified version of the Stack Focuser plugin for Fiji \cite{Umorin_2002}. The plugin works by selecting the most in focus pixel for each $(x,y)$. The most in focus pixel is the one with the highest variance of intensity in the surrounding pixels \cite{Umorin_2002}. \\

\paragraph{From experimental data to binary image}

The experimental data, the focused 2D images, was processed on the Modular Image Analysis plugin for Fiji \cite{rueden2017imagej2, schindelin2012fiji, stephen_cross_2018_2525263,Umorin_2002}. 


The image of the cell boundaries were enhanced using the WEKA pixel classification plugin, this gives us a probability image  \cite{ arganda_carreras_2016_59290}. Then a median filter is applied to the image, this removes noise from the cell centres while retaining definition of the cell-cell boundaries. Finally, cell boundaries are obtained using the watershed transform. This converts the probability image into a binary one with the boundaries labelled in black and cell in white \cite{arganda_carreras_2016_59290}. A close up of experimental data that has been processed into binary is in fig. \ref{fig:full_wing}d.
\noindent To calculate the properties of each cell they must first be approximated with a polygon. These polygons will be used for analyses later. \\

\paragraph{From binary image to polygons}

With these images, each cell was first identified and coloured. Hence, the contours between cells were able to be determined and detected. Once the contours were identified, they were fitted with polygons. Thus, the boundaries of each cell were approximated by respective polygons.

Polygons that are clearly not cells can be selectively discounted  up from the image, e.g. these could be folds in the wing where the cells are out of the frame of reference in which image slices are taken. In the supplemental information there is further information about how we identify and discount these polygons.
%
%

 An example of a polygon is shown in fig. \ref{fig:full_wing}g. This polygon approximates a cell in one of the binary images.  $(C_x,C_y)$ is the centroid (centre of mass) of the polygons. The vertices are labelled anticlockwise (this will be important in later calculations). Some notation for a polygon with $n$ vertices in calculations $\bm{r}_n = \bm{r}_0$ \cite{steger1996calculation}.

\paragraph{Mathematical tools}
We summarise here the theoretical 
tools we developed to quantify the shapes and orientations of individual cells. These local properties were used to find global averages 
and determine how they changed with time.
The area of a polygon can be obtained from taking the double integral of the domain of the shape.
\begin{align}
A &= \iint_A   \,dx\,dy \label{eq:c} 
\end{align}

\paragraph{Centroid}
The centre of the polygon (a measure of its location) is called its centroid. The centroid is the mean point of the polygon. This is where the moments of the shape are balanced.  As the polygons are 2D shapes the centroid will be a 2D vector. Each component was calculated by taking the double integral of the $x$ or $y$ over the domain of the polygon then dividing by the area.
\begin{align} 
C_x &= \frac{1}{A}\iint_A x \,dx\,dy \label{eq:a} \\
C_y &= \frac{1}{A}\iint_A y \,dx\,dy \label{eq:b}
\end{align}
%
%
\paragraph{Shape tensor}
The distribution of any molecule (e.g. a protein) in each cell (polygon) can be described by a 2nd rank tensor defined by the second moment of the area weighted with the density of that molecule. Reminiscent of the inertia tensor, it is a measure of the variance of concentration of a molecule within a cell from its centroid in different directions.  It therefore encodes information about the shape and orientation of each molecule distribution within the cell.
It is defined 
\begin{align} 
{\bm s} = \begin{pmatrix} s_{xx} & s_{xy} \\ s_{xy} & s_{yy} \end{pmatrix}
\end{align}
\noindent where
\begin{align}
s_{xx} &= -\frac{1}{A^2}\iint_A f(x', y') y'^2 \,dx'\,dy' \\
s_{xy} &= \frac{1}{A^2}\iint_A f(x', y') x' y' \,dx'\,dy' \\
s_{yy} &= -\frac{1}{A^2}\iint_A f(x', y') x'^2 \,dx'\,dy'
\end{align}
\textcolor{black}{with $y' = y - C_y$, $x' = x - C_x$ and $f(x',y')$  the concentration of said molecule. Here we have averaged over all species in the cell and taken $f(x',y') = 1$.} It is a  dimensionless quantity. Details of its computation  are given in the supplemental info.


\paragraph{Shape Factor}
From the shape tensor we can also construct a scalar, the shape factor which measures
how elongated a shape is with 0 being a round shape and approaching 1 is a very long thin shape. The shape factor is given by 
\begin{align}
s_f &= \bigg | \frac{\lambda_2 - \lambda_1}{\lambda_2 + \lambda_1} \bigg | \label{eq:Sf}
\end{align}
$s_f \in (0,1)$ as $\lambda_1, \lambda_2 > 0$. Where $\lambda_1$ and $\lambda_2$ are eigenvalues of the shape tensor $\bm s$.
%
Isotropic shapes are shapes with no clear major axis that would give it an orientation. For shapes that are isotropic their eigenvalues are a similar length making $|\lambda_1 - \lambda_2|$ small hence giving a $s_f$ close to 0. Whereas elongated shapes will have a one large eigenvalue and one small giving $|\lambda_1 - \lambda_2|$ a much larger value. 

\paragraph{Orientation}
We define the orientation as the direction of the long-axis of the polygons, as shown in fig. 1 g. The shape tensor cannot distinguish between the front or back of the polygon so the orientation is only defined modulo $\pi$. 
The orientation is defined by the eigenvalues  $\lambda_2 > \lambda_1>0$  and eigenvectors, $\bm{v}_1,\bm{v}_2$ of $\bm s$. The eigenvector corresponding to the smallest eigenvalue of $\bm s$ determines the major (long) axis and hence the orientation.
%
%
If $\bm{v}_1$ is the eigenvector of the smallest eigenvalue, 
and $\vartheta$ is the angle of orientation taken from the $x$-axis, then
\begin{align}
\vartheta &= \arctan \left(\frac{v_{12}}{v_{11}}\right) \quad \text{where} \quad \bm{v}_1 = (v_{11},v_{12})\\
\theta &= \vartheta[\pi]
\end{align}

It is more convenient for us to work with a 
q-tensor which is a traceless symmetric tensor created by 
\bem
\bmq = \bms - \mbox{Tr}(\bms) \bmI \; , 
\eem 
where $\bmI$ is the identity matrix. $\bmq = \left( \begin{array}{cc} q_1 & q_2 \\ q_2 & - q_1 \end{array} \right) \ne 0$ implies an oriented tissue and the more oriented a tissue, the greater the value of $\norm{\bmq}^2 = \frac{\sqrt{2}}{2} \mbox{Tr}(\bmq^2)$.
Now we can take the mean and standard deviation of $\bm{\hat q}_i$ for each image.

\begin{align}
    Q \bm{\hat Q} =& \frac{1}{N} \sum^N_{i=1} \bm{\hat{q}_i} , \\
    \sigma_q^2 =& \frac{1}{N} \sum^N_{i=1} \norm{\bm{\hat{q}_i} - \bm{\hat Q}}^2 \quad ,
\end{align}
where $\norm{}$ is the Frobenius norm.

\paragraph{Polarisation}
To calculate cell's polarisation, requires the use of a 3rd rank tensor. To simplify calculations we first translate and rotate the shape, $\bfr \rar \bfr'= \bfR(\theta)^{-1} \cdot \left( \bfr + \bft \right)$ such that its centroid is at the origin with its major (long) axis oriented along the $y'$-axis and the minor axis along the $x'$-axis. 
The components of the polarisation in the $x'$ and the $y'$ direction are given by $T'_{xxx}$ and $T'_{yyy}$ respectively defined below.
The polarisation vector is then defined as 
$ \quad \bmp' (\bfr') =\frac{1}{A^{5/2}} \left( T'_{xxx}, T'_{yyy}\right) \,
$, 
where
\begin{align}
    T'_{xxx} &= \iint_A x'^3 \,dx' \, dy' \\
    T'_{yyy} &= \iint_A y'^3 \,dx' \, dy' 
\end{align}
Once the polarisation vector has been calculated, the reverse rotation, $\bmp (\bfr) = \bfR (\theta) \cdot \bmp' (\bfr')$, 
gives its value in the original coordinates.
Similarly to the orientation, we can take the mean and standard deviation of $\bm{\hat p}_i$ for each image.

\begin{align}
    P \bm{\hat P} =& \frac{1}{N} \sum^N_{i=1} \bm{\hat{p}_i} , \\
    \sigma_p^2 =& \frac{1}{N} \sum^N_{i=1} (\bm{\hat{p}_i} - \bm{\hat P})^2
\end{align}

\paragraph{Theoretical model}
Given cells whose shape is given by the tensor $\bmq$, we
study the noisy growth of macroscopic orientation of the tissue along an arbitrary axis making an angle $\phi$ to the $x$-axis.  We start with a function $H (\bmq,q_0) = \sum_{i,j=1}^2 \frac{1}{2}  { \delta q_i   A_{ij}  \delta q_j }+ O(\norm{\delta \bmq}^3)$ , where $\delta \bmq = \bmq - q_0 \bmn$, which is minimised when the cell is oriented along an axis given by the tensor  
$\bmn =  \left( \begin{array}{cc} \cos 2\phi & \sin 2 \phi \\ \sin 2 \phi & - \cos 2 \phi \end{array} \right) =  \left( \begin{array}{cc} n_1 & n_2 \\ n_2 & - n_1 \end{array} \right)$. The most general form of the $2 \times 2$ matrix $A_{ij}$ which is rotationally invariant is given by $A_{ij}= A_0 \delta_{ij} + A_1 n_i n_j$ with $\delta_{ij}$ being the Kronecker delta function, and $A_{0},A_{1}$ functions of the tension in the tissue.
The 
dynamics which is a combination of gradient flow towards the minimum of $H$  and non-equilibrium stochastic driving is described by Langevin equations (Stochastic Differential Equations), for the tensor $\bmq$ and parameter $q_0$. 
\beqa
\ddt{q_i} &=& - {\partial H \over \partial q_i}  + {\xi}_i(t) = -A_{ij} \left ( q_j - q_0 n_j \right)  + {\xi}_i  \\
\ddt{q_0}  &=& k_0 - {\partial H \over \partial q_0} +  \xi_0(t)  \quad, \quad k_0 >0 \quad .
\eeqa
\tl{We emphasize that this means that the dynamics of individual cells are highly stochastic and fluctuate strongly in time.}
The fact that $k_0 >0$  implies that shape orientational order is increasing. 
$q_0(t),\bmq(t)$  have fluctuations that  are encoded in the white noises
\beq
\aver{\xi_i} = 0 \quad , \quad \aver{\xi_i (t) \xi_j (t')} = 2 D_i \delta_{ij} \delta (t-t') \quad , 
\eeq
where $i,j \in \{0,1,2\}$ and the constant parameters $k_0, A_0, A_1$ control the behaviour.  We take $D_1=D_2=D$ and $\delta(t-t')$ is Dirac delta function.
A key part of our analysis will be determining what parameters of the model are consistent with the data.

Defining $\vec X = (q_0,q_1,q_2)$
the Langevin equations for the fluctuating variables are equivalent to a Fokker-Planck equation for the probability density $P(\vec X)$ :
\beq
\partial_t P + \sum_{i=1}^3 \nabla_i  \left( V_i P\right) = 0 \quad ,  \quad V_i = ( v_i - \nabla_i H - D_i \nabla_i \ln P)
\eeq
where $\nabla_i = \partial /\partial X_i$.
The set of equations above have a steady-state probability density  given by 
\beq
P_{ss}(\bmq,A) = \frac{1}{Z} e^{-h(\bmq,A)} \quad ; \quad h = H/D_0
\eeq
and dynamical system 
\beqa
\ddt{\mean{\bmq}} &=&  - \left( 1- {D \over D_0} \right) \bmA \cdot (\mean{\bmq}-\mean{q_0} \bmn) \label{eq:dqt} \\
\ddt{\mean{q_0}} &=&  k_0 \label{eq:dq0t} \quad ,
\eeqa
as long as $D_0>D$.
It is important to note that $\min (h) = 0$ is bounded below~\cite{Liverpool2020}.

Starting with an initial typical value at $t=0$ of $\mean{q(0)}=\theta_0,\mean{\bmq(0)}=\theta_0\bmn$, we find ${q_0(t)} = \theta_0 + k_0 t, \mean{\bmq(t)}= q_0(t)\bmn$.  
This allows us to study statistically the behaviour of different trajectories (experimental samples). 
First coordinate axes are rotated so the orientation axis is along the $x$-axis.
With this we can express the steady-state probability density in terms of the deviations from the typical values. 

Hence we set $\bmq (t) = \mean{\bmq(t)}+ \delta \bmq (t), q_0 = \mean{q_0(t)} + \delta q_0 (t)$,
to obtain
\beq
P_{ss}(\delta \bmq ,  \delta q_0) = \frac{1}{Z'} \exp \left(- {1\over D_0} F (\delta q_1, \delta q_2 , \delta q_0) \right) \quad , 
\eeq
where
\beq
F =  {(A_0 + A_1) \over 2} ( \delta q_1 )^2 + {A_0 \over 2 }  ( \delta q_2 )^2 
\eeq
and $Z'$ chosen so that $\int d\bmq dq_0 P_{ss}=1$.

From this we obtain
\beqa
\aver{( \delta q_1)^2} &=& {D_0 \over A_0 + A_1}    \nonumber \\
\aver{(\delta q_2)^2 } &=& {D_0 \over A_0} \nonumber 
\eeqa


From this follows that
\beqa
\aver{\norm{\bmq}} &=& q_0(t) = \theta_0 + {k_0 t}
\eeqa

From the data if thus follows that we can extract $C,k_B,a_0,b_0,\theta$. 






\section{Results}

\paragraph{Analysis}

In this stage in development of the wing tissue, the cells are reducing their area at a linear rate. At the same time the number of cells is also increasing linearly. 
%
\paragraph{Shape Tensor}
Each cell (polygon) can be described by a 2nd rank tensor 
which we call the shape tensor $(\bmS)$, a measure of the variance of the shape of a polygon from its centroid in different directions. 
Encoding information about the shape and orientation of each cell, it is similar to the texture tensors introduced in ~\cite{Graner2008,Marmottant2008}. To compare cells of different sizes, it is made dimensionless by dividing by the area squared (see supplementary). 

\paragraph{Cell Shape Factor}
We can gain information about a cell's shape from the difference between the eigenvalues of the shape tensors, which we call the shape factor. The shape factor lies between $[0,1)$. When eigenvalues are similar this gives a low shape factor (close to 0); indicating an isotropic, non-oriented shape. A high shape factor implies an elongated shape with vastly different eigenvalues . For more details about shape tensors/factors see supplementary. 

Fig. \ref{fig:sf and ori}a, \ref{fig:sf and ori}b show the shape factor heat map and histogram respectively for a typical image.
Fig. \ref{fig:sf and ori}c shows how the average shape factor evolves with time. There is clear linear increase with time. This means cells are becoming increasingly elongated as the tissue develops. The cause of the elongation is the concentration of cells in the hinge which applies tension across the blade where we are imaging \cite{aigouy2010cell}.

\begin{figure}  [h!]
\begin{center}

	\includegraphics[scale=0.34]{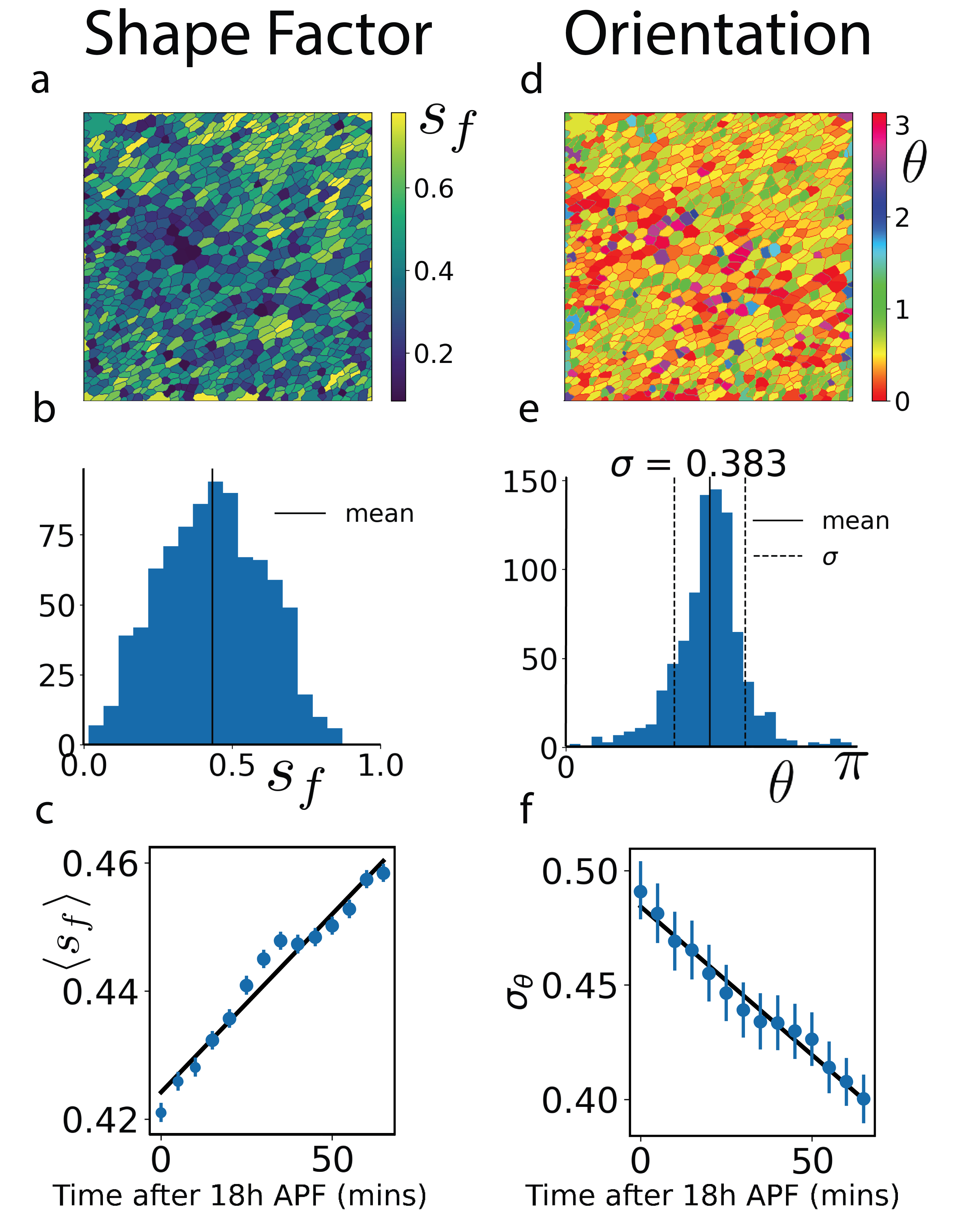} 
	\caption{\small \sl 
	a) The shape factor of cells (elongated and thin shapes have high values: yellow, isotropic (non-oriented) shapes have low values : blue) 
	b) The distribution of the shape factor. The black line shows the mean.
	c) The mean shape factor increases linearly with time. 
	d) Heat map showing  the orientation of each cell relative to the $x$-axis of a single sample.
	e) The distribution of orientations in the sample in part (c) but 'rotated' such that the mean is at $\frac{\pi}{2}$. The black line shows the mean. The standard deviation is in the subtitle.
	f) The standard deviation of the orientation averaged over 15 samples decreases linearly with time. (Errorbars are root mean square errors)}
	\label{fig:sf and ori}
\end{center}
\end{figure}

\paragraph{Cell Orientation}

We define the cell orientation as the angle, $\theta \in [0,\pi)$ between the major axis of the cell and the $x$ axis (see fig. \ref{fig:full_wing}g). 
The orientation (major) axis is the eigenvector of the shape tensor with the smallest eigenvalue. The orientation axis, invariant under reflection,  is $\pi-$periodic, $\theta = \theta + \pi$.
fig. \ref{fig:sf and ori}d shows a heatmap of the orientations from the same image as fig. \ref{fig:sf and ori}a. It is clear from the heatmap there is a global orientation of this tissue. The most common cell colour, yellow-green indicates a mean orientation of $\sim \pi/6$. 

\noindent The mean orientation,  
determined by orientation  of the pupa wing on the slide, is arbitrary. However, the standard deviation of cell orientations around the mean remains consistent between samples at the same developmental stage and indicates how oriented the tissue is; the smaller the standard deviation, the more oriented it is.
\noindent Fig. \ref{fig:sf and ori}f shows that the standard deviation is a linearly decreasing function of time, implying that an increasing number of cells are oriented in the same direction as time increases. 

We can more precisely quantify the fluctuating  growth of the tissue by studying the evolution of the whole shape tensor. 
We define a 
q-tensor, a traceless symmetric tensor created by 
\bem
\bmq = \bmS - \mbox{Tr}(\bmS) \bmI \; , 
\eem 
where $\bmI$ is the identity matrix: $\bmq = \left( \begin{array}{cc} q_1 & q_2 \\ q_2 & - q_1 \end{array} \right) = \frac{q_0}{\sqrt{2}}\left( \begin{array}{cc} \cos2\theta & \sin2\theta \\ \sin2\theta & -\cos2\theta \end{array} \right) = q_0 \hat {\bm q}$. If the norm, $\norm{\bmq} \ne 0 \quad \Rightarrow $ an elongated shape and the more oriented a shape, the greater the norm ($\norm{\bmq}^2 = \frac12 \mbox{Tr}(\bmq^2) =q^2_0$, $\hat{\bm q}$ has unit norm and $\theta$ is orientation). 

To determine the orientation of the tissue we average over q-tensors, $\bmQ=\aver{\bmq}$. We define $Q$, its magnitude (norm), and write $\bmQ=Q\hat{\bmQ}$ where $\hat{\bmQ}$ is a unit tensor.  We analyse the distribution of tensors using the standard deviation ($\sigma_q$).  If $\frac{Q}{\sigma_q} \sim 1$ then the system is functionally oriented and if $\frac{Q}{\sigma_q} \ll  1$ it is functionally  isotropic.  We find typical values of $\frac{Q}{\sigma_q} \simeq 0.84$ indicating an oriented tissue. Details about the mean and standard deviation are in the supplementary.

\paragraph{Cell Polarisation}

The shape tensor is unable to identify if a cell has any shape anisotropy along its main axes which requires a 3rd rank tensor. From this tensor we can define a polarisation vector, $\bmp$ (see supplementary).  The polarisation of a cell measures the skewness of its shape. 
%
The polarisation of each cell,  $\bmp$ can be decomposed into  magnitude  and direction : $\bmp = p \hat\bmp$, where $p=|\bmp|$ and   $\hat\bmp=(\cos \phi, \sin \phi)$. When this magnitude is large, the shape is highly polarised and when the magnitude is small,  the shape is not polarised. 
%
As above, to get an indication if the tissue has a global polarisation, it is instructive to compute the average of the polarisation vectors, $\bmP=\aver{\bmp} = P \bm{\hat P}$ and standard deviation ($\sigma_p$). $P$ is the magnitude of this average vector (how strong it is) and  the unit vector, $\bm{\hat P}$ indicates in which direction the tissue is polarised. 
If $\frac{P}{\sigma_p} \sim 1$ then the system is functionally polarised and if $\frac{P}{\sigma_p} \ll  1$ it is functionally  isotropic. Typically, values of $\frac{P}{\sigma_p} \simeq 0.021 $ indicating isotropy with no polarisation.

\paragraph{Correlations of the orientation and polarisation}

\begin{figure}  [h!]
\begin{center}

	\includegraphics[scale=0.26]{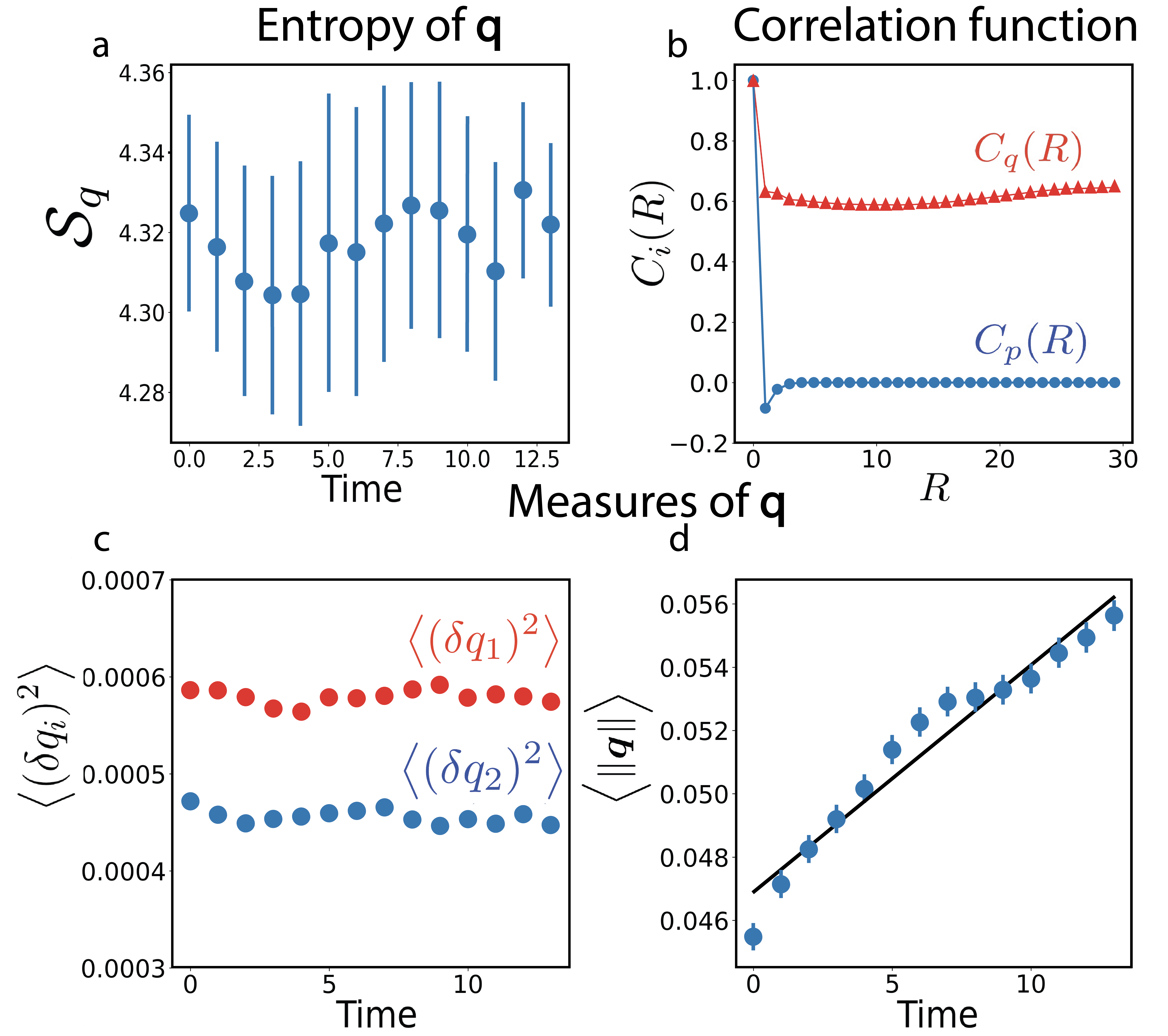} 
	\caption{\small \sl 
	a) The entropy of the $\bmq$ tenser. Each point is the mean entropy of the videos. 
	b)The figure shows the correlation function off the polarisation and the orientation respectively. Distance by a typical cell length scale is defined by the square root of the mean area of the cells in each image. $C_p(R)$ at 1 cell length starts negatively and then decays exponentially to 0. $C_q(R)$ is correlated around 0.6 throughout the tissue. (Error bars are smaller then markers).
	c) $\aver{( \delta q_1)^2}$ and $\aver{( \delta q_2)^2}$ over the frames of the video (Error bars are smaller then markers). 
	d) $\aver{\norm{\bmq}}$ over the frames of the video. This increased linearly over time and line of best fit is shown in black. (Errorbars are root mean square errors)}
	\label{fig:Q figs}
\end{center}
\end{figure}

\noindent We also consider the orientation and polarisation correlations of cells separated by distance $R$. 
\begin{align}
    C_p(R) =& \frac{1}{N_R} \sum_{R < |\bm{r_i} - \bm{r_j}| < R + dR} \langle \bm{\hat{p}_i}, \bm{\hat{p}_j}\rangle \\
    C_q(R) =& \frac{1}{N_R} \sum_{R < |\bm{r_i} - \bm{r_j}| < R + dR} \langle \bm{\hat{q}_i}, \bm{\hat{q}_j}\rangle
\end{align}
For the $\hat{\bmq}$ we use the Frobenius inner product. 
The functions, $C_p,C_q$ are plotted in fig. \ref{fig:Q figs}b. 
%
$C_p$ has a small anti-correlation at small distances after which the correlation function decays quickly to zero. It is negative at short distances which implies that if a cell is polarised in one direction its direct neighbours are likely to be polarised in the opposite direction but that cells further than one cell part are essentially uncorrelated. Therefore, while individual cells can be polarised, the tissue is not polarised at all. 
The fact that $C_q$ does not decay to zero for large distances
demonstrates that there is strong orientational order throughout the tissue. Hence the tissue is oriented but not polarised.

\paragraph{Partial Entropy production}

\tl{We also analyse the distribution of cell shapes and orientations, its evolution in time and hence extract the flow of information (partial entropy) associated with it.
At this point it is probably helpful to review some basic notions of thermodynamics and non-equilibrium statistical mechanics. We consider in general a macroscopic system plus its environment at fixed temperature which together form an isolated composite. From the 2nd law, the total entropy production is given by the sum of that produced by the system (sys) and its environment (env), \begin{math} \dot \ent_{\mbox{\footnotesize tot}}= \dot \ent_{\mbox{\footnotesize sys}}+ \dot \ent_{\mbox{\footnotesize env}} \ge  0 \; ,  \end{math}
with $\dot \ent_{{\mbox{\footnotesize sys}}} =  \dot \ent_{{\mbox{\footnotesize env}}} =0$ at equilibrium. Typically when studying passive systems, we consider situations where the system does not perturb the environment which can be considered to be at equilibrium, $\dot \ent_{\mbox{\footnotesize env}} =0$.
For a passive system coupled to such a reservoir, starting in a non-equilibrium initial state and evolving towards (a possibly more-ordered) equilibrium state, 
we would then expect that $\dot \ent_{\mbox{\footnotesize sys}} > 0 $.
We explore here what happens in a living functional tissue whose cells are becoming more orientationally ordered 
in which the system corresponds to the degrees of freedom associated with the cell shape and orientations and the environment is everything else.}

\tl{Defining the probability $\rho_{\bmq}$ of finding a cell with shape $\bmq$, we can calculate the {\em shape entropy}, the contribution to the entropy from shape fluctuations, $\ds \ent_\bmq (t)= -\sum_\bmq \rho_\bmq \ln \rho_\bmq$ and its evolution with time.  A constant value for $\ent_\bmq$ is consistent with our measurements to within the error-bars (see fig. \ref{fig:Q figs}a). Hence we find that the partial entropy production is on average zero, i.e. entropy remain constant over the whole period of observation  despite the increase in orientational order. The observation of irreversible dynamics of some observables {\em without}  increase in entropy associated with them is an indication of a flux of energy from the degrees of freedom under observation into other degrees of freedom (i.e. it implies that the environment or `reservoir' is itself not at equilibrium)~\cite{Esposito2012,Zwanzig1960}. We emphasize that this implies that while  $ \dot \ent_{\mbox{\footnotesize sys}}=0$ that $ \dot \ent_{\mbox{\footnotesize env}}>0$  and the 2nd law is still satisfied. To be concrete, the 2nd law requires that $T\dot\ent_{\mbox{\footnotesize sys}} \ge \dot E$ where $\dot E$ is the rate of energy (heat) flow to the system from the environment and $T$ is the temperature of the system (and reservoir). Since this active system has $\dot E \ne 0$, the fact that $ \dot \ent_{\mbox{\footnotesize sys}}=0$ implies that $\dot E < 0$ and there is a (steady) flow of energy (heat) to the environment from the system. Conversely the 2nd law also implies that $T\dot\ent_{\mbox{\footnotesize env}} \ge -\dot E > 0$.}

\tl{Constant entropy for a subset of the degrees of freedom, however is ideal for accurate information processing and quick response to external perturbations of those observables. This is because the accuracy threshold for any error-correction mechanism will not be changing~\cite{Voliotis2009,Mogilner2010,Sartori2013}. 
}

\paragraph{Theory}

We now develop a model of the development of tissue orientation along an axis with angle $\phi$ that can be compared quantitatively to what we observe in the experimental data.
%
The aim is to describes the dynamics of the probability  distribution of cell shapes encoded in the tensor $\bmq =  \left( \begin{array}{cc} q_1 & q_2 \\ q_2 & - q_1 \end{array} \right) $. All the data is consistent with  a steady-state probability density  given by 
\beq
P_{ss}(\bmq,q_0) = \frac{1}{Z} e^{-h(\bmq,q_0)} \quad ; \quad h = H/D_0
\label{eq:rhoss}\eeq
where $H (\bmq,q_0) = \sum_{i,j=1}^2 \frac{1}{2}  { \delta q_i   A_{ij}  \delta q_j }+ O(\norm{\delta \bmq}^3)$, where $\delta \bmq = \bmq - q_0 \bmn$ with the orientation  $\bmn =  \left( \begin{array}{cc} \cos 2\phi & \sin 2 \phi \\ \sin 2 \phi & - \cos 2 \phi \end{array} \right) =  \left( \begin{array}{cc} n_1 & n_2 \\ n_2 & - n_1 \end{array} \right)$. Note that $\mbox{min}(h) > -\infty$, must be bounded for eqn.(\ref{eq:rhoss}) to make sense. The most general form of the matrix $\bmA$ which is rotationally invariant is given by $A_{ij}= A_0 \delta_{ij} + A_1 n_i n_j$. The {\em fluctuating} variables $q_0(t),q_1(t),q_2(t)$ capture the changes in cell shape which evolve {\em on average} according to the deterministic dynamical system 
\beqa
\ddt{\mean{\bmq}} &=& {\bm{V}}(\mean{\bmq},\mean{q_0})= - \bmA \cdot (\mean{\bmq}-\mean{q_0} \bmn) \label{eq:dqt} \\
\ddt{\mean{q_0}} &=& {{V}_0}(\mean{\bmq},\mean{q_0})=  k_0 \label{eq:dq0t}
\eeqa
\tl{It is helpful at this point to compare this with the stochastic dynamics of an equilibrium system with the same number of degrees of freedom $\bm{e}(t)=\left(e_0,e_1,e_2\right)$ and a Hamitonian $\ham (\bm{e})$ so that \beq
\ddt{{e}_i} = - {\partial \ham \over \partial e_i} + {\xi}_i(t) \quad , \quad 
\eeq
\beq
\aver{\xi_i} = 0 \quad , \quad \aver{\xi_i (t) \xi_j (t')} = 2 T \delta_{ij} \delta (t-t') \quad , 
\eeq
This system has a steady-state with density of states, $P_{eq}(\bm{e})=e^{-\ham/T}$ with average constant values $\mean{e_i}$ which means in the steady state the average, $\mean{\bm{e}(t)}$ evolves according to 
\beq
\ddt{\mean{e_i}} = 0 \quad , 
\eeq
i.e. the d.o.f do not change with time.  The dynamical system of the average behaviour is a trivial one: all velocities are on average zero and each realisation  of the system (or equivalently each experiment) simply fluctuates around the mean constant value. This is equivalent to a detailed balance condition that is required by all systems at equilibrium.
Here, the existence of a non-trivial dynamical system (i.e. the fact that the rhs (right hand side) of eqns. (\ref{eq:dqt},\ref{eq:dq0t}) are not zero) means that the living epithelium cannot be mapped to an equilibrium system. Consequently detailed balance is also broken:  there are non-zero currents $\vec J_{ss}(\bmq,q_0)  = \left(\bm{V} (\bmq,q_0), V_0 (\bmq,q_0) \right) P_{ss}(\bmq,q_0)$ which break detailed balance. Each realisation (experiment) will fluctuate around these deterministic trajectories.}
Guided by the data, we will obtain values for $A_0,A_1,k_0$.
This allows us to study statistically the behaviour of different trajectories (experimental samples). 
First coordinate axes are rotated so the orientation axis is along the $x$-axis (i.e. $n_1=1,n_2=0$).
With this we can express the steady-state probability density in terms of the deviations from the typical values $\mean{q_0(t)},\mean{q_1 (t)},\mean{q_2(t)}$.
%
%
From the data we can compute $\aver{\norm{\bmq}}$, $\aver{( \delta q_1)^2}$ and $\aver{( \delta q_2)^2}$, this is shown in fig. \ref{fig:Q figs}c-d, from which  we can extract $k_0,A_0,A_1$ (see supplementary)   through it is not clear how these relate to mechanical properties of the developing tissue. 

\section{Summary and Discussion}

\tl{The ability to visualise the dynamic evolution of the spatial distribution of specific proteins within individual cells in tissue provided by high resolution imaging  holds promise for us eventually being able to extract the organising principles behind tissue function and repair. This biological function happens in the presence of large fluctuations, both chemical and mechanical, hence  these principles, whatever they turn out to be, must be robust to noise.  It is the implications of this robustness that we focus on in this paper. Here we have tracked the concentration of junction proteins to  quantify the dynamics of  cell shapes, orientations and polarisation measuring not only their averages but most importantly their fluctuations. 
In this context, the degrees of freedom associated with the shapes, orientations and polarisations of the cells are the {\em system} while everything else in the tissue we consider as the {\em environment}.
\\
The experimental data in total paint a consistent picture. We observe irreversible dynamics of the cells becoming more elongated and the tissue 
becoming more oriented along a particular direction $\bfn$ without any associated change in entropy. We emphasize that this implies that while the rate of entropy production of the environment is non-zero,  $ \dot \ent_{\mbox{\footnotesize env}}>0$ that the rate of change of entropy associated with those degrees of freedom in the tissue, $ \dot \ent_{\mbox{\footnotesize sys}}=0$. This implies a steady flow of energy from the reservoir to the system. Given that the reservoir is much larger than the system, $ \ent_{\mbox{\footnotesize sys}} \ll   \ent_{\mbox{\footnotesize env}}$, we note that it is also possible that at different points in development, 
the rate of change of entropy in the tissue, $ \dot \ent_{\mbox{\footnotesize sys}}$ can be non-zero,  even negative. 
}

We also find no macroscopic shape polarisation of the tissue; i.e. we find nematic symmetry of the director axis, i.e. $\bfn$ and $-\bfn$ are equivalent.  
This can be explained by a model of cell shape and orientation that can be mapped to a  non-equilibrium driven nematic liquid crystal. 
We find that while oriented, the tissues have no global polarisation and that cells that are polarised only affect other cells in their very close proximity.
An observation from our data is that all these features are developed  in the presence of strong local variation and fluctuations  about the average behaviour. This indicates a large but constant information entropy. \tl{Clearly this approach can be extended to any group of observables in living tissue or other functional biological matter.} This suggests an intriguing possibility that should be investigated further  by studying the statistical dynamics of other chemical and geometrical quantities in tissue to see if  zero rate of change of entropy with irreversible dynamics is a signature of functionality and homeostasis in healthy living organisms.
This characterisation of an unperturbed but dynamic, developing tissue will lay the groundwork for understanding what happens when these tissues are perturbed for example by cancer or wounding.

{The computational resources of the University of Bristol Advanced Computing Research Centre, and the BrisSynBio HPC facility are gratefully acknowledged. MO acknowledges the support of the Wellcome Trust. JT acknowledges a MRC GW4 studentship. TBL acknowledges support of Leverhulme Trust Research Project Grant RPG-2016-147 and BrisSynBio, a BBSRC/EPSRC Synthetic Biology Research Center (BB/L01386X/1). }

\medskip

\bibliographystyle{apsrev4-1}
\bibliography{main} 

\begin{thebibliography}{52}%
\makeatletter
\providecommand \@ifxundefined [1]{%
 \@ifx{#1\undefined}
}%
\providecommand \@ifnum [1]{%
 \ifnum #1\expandafter \@firstoftwo
 \else \expandafter \@secondoftwo
 \fi
}%
\providecommand \@ifx [1]{%
 \ifx #1\expandafter \@firstoftwo
 \else \expandafter \@secondoftwo
 \fi
}%
\providecommand \natexlab [1]{#1}%
\providecommand \enquote  [1]{``#1''}%
\providecommand \bibnamefont  [1]{#1}%
\providecommand \bibfnamefont [1]{#1}%
\providecommand \citenamefont [1]{#1}%
\providecommand \href@noop [0]{\@secondoftwo}%
\providecommand \href [0]{\begingroup \@sanitize@url \@href}%
\providecommand \@href[1]{\@@startlink{#1}\@@href}%
\providecommand \@@href[1]{\endgroup#1\@@endlink}%
\providecommand \@sanitize@url [0]{\catcode `\\12\catcode `\$12\catcode
  `\&12\catcode `\#12\catcode `\^12\catcode `\_12\catcode `\%12\relax}%
\providecommand \@@startlink[1]{}%
\providecommand \@@endlink[0]{}%
\providecommand \url  [0]{\begingroup\@sanitize@url \@url }%
\providecommand \@url [1]{\endgroup\@href {#1}{\urlprefix }}%
\providecommand \urlprefix  [0]{URL }%
\providecommand \Eprint [0]{\href }%
\providecommand \doibase [0]{http://dx.doi.org/}%
\providecommand \selectlanguage [0]{\@gobble}%
\providecommand \bibinfo  [0]{\@secondoftwo}%
\providecommand \bibfield  [0]{\@secondoftwo}%
\providecommand \translation [1]{[#1]}%
\providecommand \BibitemOpen [0]{}%
\providecommand \bibitemStop [0]{}%
\providecommand \bibitemNoStop [0]{.\EOS\space}%
\providecommand \EOS [0]{\spacefactor3000\relax}%
\providecommand \BibitemShut  [1]{\csname bibitem#1\endcsname}%
\let\auto@bib@innerbib\@empty
\bibitem [{\citenamefont {Weavers}\ \emph {et~al.}(2018)\citenamefont
  {Weavers}, \citenamefont {Franz}, \citenamefont {Wood},\ and\ \citenamefont
  {Martin}}]{weavers2018long}%
  \BibitemOpen
  \bibfield  {author} {\bibinfo {author} {\bibfnamefont {H.}~\bibnamefont
  {Weavers}}, \bibinfo {author} {\bibfnamefont {A.}~\bibnamefont {Franz}},
  \bibinfo {author} {\bibfnamefont {W.}~\bibnamefont {Wood}}, \ and\ \bibinfo
  {author} {\bibfnamefont {P.}~\bibnamefont {Martin}},\ }\href@noop {}
  {\bibfield  {journal} {\bibinfo  {journal} {Journal of visualized
  experiments: JoVE}\ } (\bibinfo {year} {2018})}\BibitemShut {NoStop}%
\bibitem [{\citenamefont {Anderson}(1972)}]{Andersen1972}%
  \BibitemOpen
  \bibfield  {author} {\bibinfo {author} {\bibfnamefont {P.~W.}\ \bibnamefont
  {Anderson}},\ }\href {\doibase 10.1126/science.177.4047.393} {\bibfield
  {journal} {\bibinfo  {journal} {Science}\ }\textbf {\bibinfo {volume}
  {177}},\ \bibinfo {pages} {393} (\bibinfo {year} {1972})}\BibitemShut
  {NoStop}%
\bibitem [{\citenamefont {Lecuit}\ and\ \citenamefont
  {Lenne}(2007)}]{Lecuit2007a}%
  \BibitemOpen
  \bibfield  {author} {\bibinfo {author} {\bibfnamefont {T.}~\bibnamefont
  {Lecuit}}\ and\ \bibinfo {author} {\bibfnamefont {P.~F.}\ \bibnamefont
  {Lenne}},\ }\href {\doibase 10.1038/nrm2222} {\bibfield  {journal} {\bibinfo
  {journal} {Nature Reviews Molecular Cell Biology}\ }\textbf {\bibinfo
  {volume} {8}},\ \bibinfo {pages} {633} (\bibinfo {year} {2007})}\BibitemShut
  {NoStop}%
\bibitem [{\citenamefont {{Le Goff}}\ and\ \citenamefont
  {Lecuit}(2011)}]{Lecuit2011}%
  \BibitemOpen
  \bibfield  {author} {\bibinfo {author} {\bibfnamefont {L.}~\bibnamefont {{Le
  Goff}}}\ and\ \bibinfo {author} {\bibfnamefont {T.}~\bibnamefont {Lecuit}},\
  }\href@noop {} {\bibfield  {journal} {\bibinfo  {journal} {Science}\ }\textbf
  {\bibinfo {volume} {331}},\ \bibinfo {pages} {1141} (\bibinfo {year}
  {2011})}\BibitemShut {NoStop}%
\bibitem [{\citenamefont {Martin}\ and\ \citenamefont
  {Lewis}(1992)}]{Martin1992}%
  \BibitemOpen
  \bibfield  {author} {\bibinfo {author} {\bibfnamefont {P.}~\bibnamefont
  {Martin}}\ and\ \bibinfo {author} {\bibfnamefont {J.}~\bibnamefont {Lewis}},\
  }\href {\doibase 10.1038/360179a0} {\bibfield  {journal} {\bibinfo  {journal}
  {Nature}\ }\textbf {\bibinfo {volume} {360}},\ \bibinfo {pages} {179}
  (\bibinfo {year} {1992})}\BibitemShut {NoStop}%
\bibitem [{\citenamefont {Razzell}\ \emph {et~al.}(2014)\citenamefont
  {Razzell}, \citenamefont {Wood},\ and\ \citenamefont
  {Martin}}]{Razzell2014a}%
  \BibitemOpen
  \bibfield  {author} {\bibinfo {author} {\bibfnamefont {W.}~\bibnamefont
  {Razzell}}, \bibinfo {author} {\bibfnamefont {W.}~\bibnamefont {Wood}}, \
  and\ \bibinfo {author} {\bibfnamefont {P.}~\bibnamefont {Martin}},\ }\href
  {\doibase 10.1242/dev.107045} {\bibfield  {journal} {\bibinfo  {journal}
  {Development}\ }\textbf {\bibinfo {volume} {141}},\ \bibinfo {pages} {1814}
  (\bibinfo {year} {2014})}\BibitemShut {NoStop}%
\bibitem [{\citenamefont {Tetley}\ \emph {et~al.}(2019)\citenamefont {Tetley},
  \citenamefont {Staddon}, \citenamefont {Heller}, \citenamefont {Hoppe},
  \citenamefont {Banerjee},\ and\ \citenamefont {Mao}}]{Tetley2019}%
  \BibitemOpen
  \bibfield  {author} {\bibinfo {author} {\bibfnamefont {R.~J.}\ \bibnamefont
  {Tetley}}, \bibinfo {author} {\bibfnamefont {M.~F.}\ \bibnamefont {Staddon}},
  \bibinfo {author} {\bibfnamefont {D.}~\bibnamefont {Heller}}, \bibinfo
  {author} {\bibfnamefont {A.}~\bibnamefont {Hoppe}}, \bibinfo {author}
  {\bibfnamefont {S.}~\bibnamefont {Banerjee}}, \ and\ \bibinfo {author}
  {\bibfnamefont {Y.}~\bibnamefont {Mao}},\ }\href {\doibase
  10.1038/s41567-019-0618-1} {\bibfield  {journal} {\bibinfo  {journal} {Nature
  Physics}\ }\textbf {\bibinfo {volume} {15}},\ \bibinfo {pages} {1195}
  (\bibinfo {year} {2019})}\BibitemShut {NoStop}%
\bibitem [{\citenamefont {Turley}\ \emph {et~al.}(2022)\citenamefont {Turley},
  \citenamefont {Chenchiah}, \citenamefont {Liverpool}, \citenamefont
  {Weavers},\ and\ \citenamefont {Martin}}]{turley2022good}%
  \BibitemOpen
  \bibfield  {author} {\bibinfo {author} {\bibfnamefont {J.}~\bibnamefont
  {Turley}}, \bibinfo {author} {\bibfnamefont {I.~V.}\ \bibnamefont
  {Chenchiah}}, \bibinfo {author} {\bibfnamefont {T.~B.}\ \bibnamefont
  {Liverpool}}, \bibinfo {author} {\bibfnamefont {H.}~\bibnamefont {Weavers}},
  \ and\ \bibinfo {author} {\bibfnamefont {P.}~\bibnamefont {Martin}},\
  }\href@noop {} {\bibfield  {journal} {\bibinfo  {journal} {Iscience}\ ,\
  \bibinfo {pages} {104778}} (\bibinfo {year} {2022})}\BibitemShut {NoStop}%
\bibitem [{\citenamefont {Piersma}\ \emph {et~al.}(2020)\citenamefont
  {Piersma}, \citenamefont {Hayward},\ and\ \citenamefont
  {Weaver}}]{piersma2020fibrosis}%
  \BibitemOpen
  \bibfield  {author} {\bibinfo {author} {\bibfnamefont {B.}~\bibnamefont
  {Piersma}}, \bibinfo {author} {\bibfnamefont {M.}~\bibnamefont {Hayward}}, \
  and\ \bibinfo {author} {\bibfnamefont {V.~M.}\ \bibnamefont {Weaver}},\
  }\href@noop {} {\bibfield  {journal} {\bibinfo  {journal} {Biochimica et
  Biophysica Acta (BBA)-Reviews on Cancer}\ }\textbf {\bibinfo {volume}
  {1873}},\ \bibinfo {pages} {188356} (\bibinfo {year} {2020})}\BibitemShut
  {NoStop}%
\bibitem [{\citenamefont {Hutson}\ \emph {et~al.}(2003)\citenamefont {Hutson},
  \citenamefont {Tokutake}, \citenamefont {Chang}, \citenamefont {Bloor},
  \citenamefont {Venakides}, \citenamefont {Kiehart},\ and\ \citenamefont
  {Edwards}}]{hutson2003forces}%
  \BibitemOpen
  \bibfield  {author} {\bibinfo {author} {\bibfnamefont {M.~S.}\ \bibnamefont
  {Hutson}}, \bibinfo {author} {\bibfnamefont {Y.}~\bibnamefont {Tokutake}},
  \bibinfo {author} {\bibfnamefont {M.-S.}\ \bibnamefont {Chang}}, \bibinfo
  {author} {\bibfnamefont {J.~W.}\ \bibnamefont {Bloor}}, \bibinfo {author}
  {\bibfnamefont {S.}~\bibnamefont {Venakides}}, \bibinfo {author}
  {\bibfnamefont {D.~P.}\ \bibnamefont {Kiehart}}, \ and\ \bibinfo {author}
  {\bibfnamefont {G.~S.}\ \bibnamefont {Edwards}},\ }\href@noop {} {\bibfield
  {journal} {\bibinfo  {journal} {Science}\ }\textbf {\bibinfo {volume}
  {300}},\ \bibinfo {pages} {145} (\bibinfo {year} {2003})}\BibitemShut
  {NoStop}%
\bibitem [{\citenamefont {Ishihara}\ and\ \citenamefont
  {Sugimura}(2012)}]{ishihara2012bayesian}%
  \BibitemOpen
  \bibfield  {author} {\bibinfo {author} {\bibfnamefont {S.}~\bibnamefont
  {Ishihara}}\ and\ \bibinfo {author} {\bibfnamefont {K.}~\bibnamefont
  {Sugimura}},\ }\href@noop {} {\bibfield  {journal} {\bibinfo  {journal}
  {Journal of theoretical biology}\ }\textbf {\bibinfo {volume} {313}},\
  \bibinfo {pages} {201} (\bibinfo {year} {2012})}\BibitemShut {NoStop}%
\bibitem [{\citenamefont {Sugimura}\ and\ \citenamefont
  {Ishihara}(2013)}]{sugimura2013mechanical}%
  \BibitemOpen
  \bibfield  {author} {\bibinfo {author} {\bibfnamefont {K.}~\bibnamefont
  {Sugimura}}\ and\ \bibinfo {author} {\bibfnamefont {S.}~\bibnamefont
  {Ishihara}},\ }\href@noop {} {\bibfield  {journal} {\bibinfo  {journal}
  {Development}\ }\textbf {\bibinfo {volume} {140}},\ \bibinfo {pages} {4091}
  (\bibinfo {year} {2013})}\BibitemShut {NoStop}%
\bibitem [{\citenamefont {Etournay}\ \emph {et~al.}(2015)\citenamefont
  {Etournay}, \citenamefont {Popovi{\'c}}, \citenamefont {Merkel},
  \citenamefont {Nandi}, \citenamefont {Blasse}, \citenamefont {Aigouy},
  \citenamefont {Brandl}, \citenamefont {Myers}, \citenamefont {Salbreux},
  \citenamefont {J{\"u}licher} \emph {et~al.}}]{etournay2015interplay}%
  \BibitemOpen
  \bibfield  {author} {\bibinfo {author} {\bibfnamefont {R.}~\bibnamefont
  {Etournay}}, \bibinfo {author} {\bibfnamefont {M.}~\bibnamefont
  {Popovi{\'c}}}, \bibinfo {author} {\bibfnamefont {M.}~\bibnamefont {Merkel}},
  \bibinfo {author} {\bibfnamefont {A.}~\bibnamefont {Nandi}}, \bibinfo
  {author} {\bibfnamefont {C.}~\bibnamefont {Blasse}}, \bibinfo {author}
  {\bibfnamefont {B.}~\bibnamefont {Aigouy}}, \bibinfo {author} {\bibfnamefont
  {H.}~\bibnamefont {Brandl}}, \bibinfo {author} {\bibfnamefont
  {G.}~\bibnamefont {Myers}}, \bibinfo {author} {\bibfnamefont
  {G.}~\bibnamefont {Salbreux}}, \bibinfo {author} {\bibfnamefont
  {F.}~\bibnamefont {J{\"u}licher}},  \emph {et~al.},\ }\href@noop {}
  {\bibfield  {journal} {\bibinfo  {journal} {Elife}\ }\textbf {\bibinfo
  {volume} {4}},\ \bibinfo {pages} {e07090} (\bibinfo {year}
  {2015})}\BibitemShut {NoStop}%
\bibitem [{\citenamefont {Elowitz}\ \emph {et~al.}(2002)\citenamefont
  {Elowitz}, \citenamefont {Levine}, \citenamefont {Siggia},\ and\
  \citenamefont {Swain}}]{Elowitz2002}%
  \BibitemOpen
  \bibfield  {author} {\bibinfo {author} {\bibfnamefont {M.~B.}\ \bibnamefont
  {Elowitz}}, \bibinfo {author} {\bibfnamefont {A.~J.}\ \bibnamefont {Levine}},
  \bibinfo {author} {\bibfnamefont {E.~D.}\ \bibnamefont {Siggia}}, \ and\
  \bibinfo {author} {\bibfnamefont {P.~S.}\ \bibnamefont {Swain}},\ }\href
  {\doibase 10.1126/science.1070919} {\bibfield  {journal} {\bibinfo  {journal}
  {Science (New York, N.Y.)}\ }\textbf {\bibinfo {volume} {297}},\ \bibinfo
  {pages} {1183} (\bibinfo {year} {2002})}\BibitemShut {NoStop}%
\bibitem [{\citenamefont {Chubb}\ and\ \citenamefont
  {Liverpool}(2010)}]{Chubb2010}%
  \BibitemOpen
  \bibfield  {author} {\bibinfo {author} {\bibfnamefont {J.~R.}\ \bibnamefont
  {Chubb}}\ and\ \bibinfo {author} {\bibfnamefont {T.~B.}\ \bibnamefont
  {Liverpool}},\ }\href {\doibase 10.1016/j.gde.2010.06.009} {\bibfield
  {journal} {\bibinfo  {journal} {Current opinion in genetics {\&}
  development}\ }\textbf {\bibinfo {volume} {20}},\ \bibinfo {pages} {478}
  (\bibinfo {year} {2010})}\BibitemShut {NoStop}%
\bibitem [{\citenamefont {Aigouy}\ \emph
  {et~al.}(2010{\natexlab{a}})\citenamefont {Aigouy}, \citenamefont
  {Farhadifar}, \citenamefont {Staple}, \citenamefont {Sagner}, \citenamefont
  {R{\"{o}}per}, \citenamefont {J{\"{u}}licher},\ and\ \citenamefont
  {Eaton}}]{Aigouy2010a}%
  \BibitemOpen
  \bibfield  {author} {\bibinfo {author} {\bibfnamefont {B.}~\bibnamefont
  {Aigouy}}, \bibinfo {author} {\bibfnamefont {R.}~\bibnamefont {Farhadifar}},
  \bibinfo {author} {\bibfnamefont {D.~B.}\ \bibnamefont {Staple}}, \bibinfo
  {author} {\bibfnamefont {A.}~\bibnamefont {Sagner}}, \bibinfo {author}
  {\bibfnamefont {J.-C.}\ \bibnamefont {R{\"{o}}per}}, \bibinfo {author}
  {\bibfnamefont {F.}~\bibnamefont {J{\"{u}}licher}}, \ and\ \bibinfo {author}
  {\bibfnamefont {S.}~\bibnamefont {Eaton}},\ }\href {\doibase
  10.1016/j.cell.2010.07.042} {\bibfield  {journal} {\bibinfo  {journal}
  {Cell}\ }\textbf {\bibinfo {volume} {142}},\ \bibinfo {pages} {773} (\bibinfo
  {year} {2010}{\natexlab{a}})}\BibitemShut {NoStop}%
\bibitem [{\citenamefont {Duclos}\ \emph {et~al.}(2014)\citenamefont {Duclos},
  \citenamefont {Garcia}, \citenamefont {Yevick},\ and\ \citenamefont
  {Silberzan}}]{Duclos2014}%
  \BibitemOpen
  \bibfield  {author} {\bibinfo {author} {\bibfnamefont {G.}~\bibnamefont
  {Duclos}}, \bibinfo {author} {\bibfnamefont {S.}~\bibnamefont {Garcia}},
  \bibinfo {author} {\bibfnamefont {H.~G.}\ \bibnamefont {Yevick}}, \ and\
  \bibinfo {author} {\bibfnamefont {P.}~\bibnamefont {Silberzan}},\ }\href
  {\doibase 10.1039/C3SM52323C} {\bibfield  {journal} {\bibinfo  {journal}
  {Soft Matter}\ }\textbf {\bibinfo {volume} {10}},\ \bibinfo {pages} {2346}
  (\bibinfo {year} {2014})}\BibitemShut {NoStop}%
\bibitem [{\citenamefont {Duclos}\ \emph {et~al.}(2017)\citenamefont {Duclos},
  \citenamefont {Erlenk{\"{a}}mper}, \citenamefont {Joanny},\ and\
  \citenamefont {Silberzan}}]{Duclos2017b}%
  \BibitemOpen
  \bibfield  {author} {\bibinfo {author} {\bibfnamefont {G.}~\bibnamefont
  {Duclos}}, \bibinfo {author} {\bibfnamefont {C.}~\bibnamefont
  {Erlenk{\"{a}}mper}}, \bibinfo {author} {\bibfnamefont {J.-F.}\ \bibnamefont
  {Joanny}}, \ and\ \bibinfo {author} {\bibfnamefont {P.}~\bibnamefont
  {Silberzan}},\ }\href {\doibase 10.1038/nphys3876} {\bibfield  {journal}
  {\bibinfo  {journal} {Nature Physics}\ }\textbf {\bibinfo {volume} {13}},\
  \bibinfo {pages} {58} (\bibinfo {year} {2017})}\BibitemShut {NoStop}%
\bibitem [{\citenamefont {Li}\ \emph {et~al.}(2017)\citenamefont {Li},
  \citenamefont {Balagam}, \citenamefont {He}, \citenamefont {Lee},
  \citenamefont {Igoshin},\ and\ \citenamefont {Levine}}]{Li2017}%
  \BibitemOpen
  \bibfield  {author} {\bibinfo {author} {\bibfnamefont {X.}~\bibnamefont
  {Li}}, \bibinfo {author} {\bibfnamefont {R.}~\bibnamefont {Balagam}},
  \bibinfo {author} {\bibfnamefont {T.-F.}\ \bibnamefont {He}}, \bibinfo
  {author} {\bibfnamefont {P.~P.}\ \bibnamefont {Lee}}, \bibinfo {author}
  {\bibfnamefont {O.~A.}\ \bibnamefont {Igoshin}}, \ and\ \bibinfo {author}
  {\bibfnamefont {H.}~\bibnamefont {Levine}},\ }\href {\doibase
  10.1073/pnas.1707210114} {\bibfield  {journal} {\bibinfo  {journal}
  {Proceedings of the National Academy of Sciences}\ }\textbf {\bibinfo
  {volume} {114}},\ \bibinfo {pages} {8974} (\bibinfo {year}
  {2017})}\BibitemShut {NoStop}%
\bibitem [{\citenamefont {Saw}\ \emph {et~al.}(2017)\citenamefont {Saw},
  \citenamefont {Doostmohammadi}, \citenamefont {Nier}, \citenamefont
  {Kocgozlu}, \citenamefont {Thampi}, \citenamefont {Toyama}, \citenamefont
  {Marcq}, \citenamefont {Lim}, \citenamefont {Yeomans},\ and\ \citenamefont
  {Ladoux}}]{Saw2017}%
  \BibitemOpen
  \bibfield  {author} {\bibinfo {author} {\bibfnamefont {T.~B.}\ \bibnamefont
  {Saw}}, \bibinfo {author} {\bibfnamefont {A.}~\bibnamefont {Doostmohammadi}},
  \bibinfo {author} {\bibfnamefont {V.}~\bibnamefont {Nier}}, \bibinfo {author}
  {\bibfnamefont {L.}~\bibnamefont {Kocgozlu}}, \bibinfo {author}
  {\bibfnamefont {S.}~\bibnamefont {Thampi}}, \bibinfo {author} {\bibfnamefont
  {Y.}~\bibnamefont {Toyama}}, \bibinfo {author} {\bibfnamefont
  {P.}~\bibnamefont {Marcq}}, \bibinfo {author} {\bibfnamefont {C.~T.}\
  \bibnamefont {Lim}}, \bibinfo {author} {\bibfnamefont {J.~M.}\ \bibnamefont
  {Yeomans}}, \ and\ \bibinfo {author} {\bibfnamefont {B.}~\bibnamefont
  {Ladoux}},\ }\href {\doibase 10.1038/nature21718} {\bibfield  {journal}
  {\bibinfo  {journal} {Nature}\ }\textbf {\bibinfo {volume} {544}},\ \bibinfo
  {pages} {212} (\bibinfo {year} {2017})}\BibitemShut {NoStop}%
\bibitem [{\citenamefont {Kawaguchi}\ \emph {et~al.}(2017)\citenamefont
  {Kawaguchi}, \citenamefont {Kageyama},\ and\ \citenamefont
  {Sano}}]{Kawaguchi2017}%
  \BibitemOpen
  \bibfield  {author} {\bibinfo {author} {\bibfnamefont {K.}~\bibnamefont
  {Kawaguchi}}, \bibinfo {author} {\bibfnamefont {R.}~\bibnamefont {Kageyama}},
  \ and\ \bibinfo {author} {\bibfnamefont {M.}~\bibnamefont {Sano}},\ }\href
  {\doibase 10.1038/nature22321} {\bibfield  {journal} {\bibinfo  {journal}
  {Nature}\ }\textbf {\bibinfo {volume} {545}},\ \bibinfo {pages} {327}
  (\bibinfo {year} {2017})}\BibitemShut {NoStop}%
\bibitem [{\citenamefont {Martella}\ \emph {et~al.}(2019)\citenamefont
  {Martella}, \citenamefont {Pattelli}, \citenamefont {Matassini},
  \citenamefont {Ridi}, \citenamefont {Bonini}, \citenamefont {Paoli},
  \citenamefont {Baglioni}, \citenamefont {Wiersma},\ and\ \citenamefont
  {Parmeggiani}}]{Martella2019}%
  \BibitemOpen
  \bibfield  {author} {\bibinfo {author} {\bibfnamefont {D.}~\bibnamefont
  {Martella}}, \bibinfo {author} {\bibfnamefont {L.}~\bibnamefont {Pattelli}},
  \bibinfo {author} {\bibfnamefont {C.}~\bibnamefont {Matassini}}, \bibinfo
  {author} {\bibfnamefont {F.}~\bibnamefont {Ridi}}, \bibinfo {author}
  {\bibfnamefont {M.}~\bibnamefont {Bonini}}, \bibinfo {author} {\bibfnamefont
  {P.}~\bibnamefont {Paoli}}, \bibinfo {author} {\bibfnamefont
  {P.}~\bibnamefont {Baglioni}}, \bibinfo {author} {\bibfnamefont {D.~S.}\
  \bibnamefont {Wiersma}}, \ and\ \bibinfo {author} {\bibfnamefont
  {C.}~\bibnamefont {Parmeggiani}},\ }\href {\doibase 10.1002/adhm.201801489}
  {\bibfield  {journal} {\bibinfo  {journal} {Advanced Healthcare Materials}\
  }\textbf {\bibinfo {volume} {8}},\ \bibinfo {pages} {1801489} (\bibinfo
  {year} {2019})}\BibitemShut {NoStop}%
\bibitem [{\citenamefont {Brenner}(2012)}]{Brenner2012}%
  \BibitemOpen
  \bibfield  {author} {\bibinfo {author} {\bibfnamefont {S.}~\bibnamefont
  {Brenner}},\ }\href {\doibase 10.1038/482461a} {\bibfield  {journal}
  {\bibinfo  {journal} {Nature}\ }\textbf {\bibinfo {volume} {482}},\ \bibinfo
  {pages} {461} (\bibinfo {year} {2012})}\BibitemShut {NoStop}%
\bibitem [{\citenamefont {Ahmadzade}\ \emph {et~al.}(2017)\citenamefont
  {Ahmadzade}, \citenamefont {Gao}, \citenamefont {Dehghan},\ and\
  \citenamefont {Sheng}}]{Ahmadzade2017}%
  \BibitemOpen
  \bibfield  {author} {\bibinfo {author} {\bibfnamefont {H.}~\bibnamefont
  {Ahmadzade}}, \bibinfo {author} {\bibfnamefont {R.}~\bibnamefont {Gao}},
  \bibinfo {author} {\bibfnamefont {M.~H.}\ \bibnamefont {Dehghan}}, \ and\
  \bibinfo {author} {\bibfnamefont {Y.}~\bibnamefont {Sheng}},\ }\href@noop {}
  {\bibfield  {journal} {\bibinfo  {journal} {Journal of Intelligent \& Fuzzy
  Systems}\ }\textbf {\bibinfo {volume} {33}},\ \bibinfo {pages} {105}
  (\bibinfo {year} {2017})}\BibitemShut {NoStop}%
\bibitem [{\citenamefont {Bouligand}(2008)}]{Bouligand2008}%
  \BibitemOpen
  \bibfield  {author} {\bibinfo {author} {\bibfnamefont {Y.}~\bibnamefont
  {Bouligand}},\ }\href {\doibase 10.1016/j.crci.2007.10.001} {\bibfield
  {journal} {\bibinfo  {journal} {Comptes Rendus Chimie}\ }\textbf {\bibinfo
  {volume} {11}},\ \bibinfo {pages} {281} (\bibinfo {year} {2008})}\BibitemShut
  {NoStop}%
\bibitem [{\citenamefont {Gennes}\ and\ \citenamefont
  {Prost}(1993)}]{DeGennesProst93}%
  \BibitemOpen
  \bibfield  {author} {\bibinfo {author} {\bibfnamefont {P.~G.~D.}\
  \bibnamefont {Gennes}}\ and\ \bibinfo {author} {\bibfnamefont
  {J.}~\bibnamefont {Prost}},\ }\href@noop {} {\emph {\bibinfo {title} {{The
  Physics of Liquid Crystals}}}}\ (\bibinfo  {publisher} {Clarendon Press},\
  \bibinfo {address} {Oxford},\ \bibinfo {year} {1993})\BibitemShut {NoStop}%
\bibitem [{\citenamefont {Duvert}\ \emph {et~al.}(1984)\citenamefont {Duvert},
  \citenamefont {Bouligand},\ and\ \citenamefont {Salat}}]{Duvert1984}%
  \BibitemOpen
  \bibfield  {author} {\bibinfo {author} {\bibfnamefont {M.}~\bibnamefont
  {Duvert}}, \bibinfo {author} {\bibfnamefont {Y.}~\bibnamefont {Bouligand}}, \
  and\ \bibinfo {author} {\bibfnamefont {C.}~\bibnamefont {Salat}},\ }\href
  {\doibase 10.1016/0040-8166(84)90064-8} {\bibfield  {journal} {\bibinfo
  {journal} {Tissue and Cell}\ }\textbf {\bibinfo {volume} {16}},\ \bibinfo
  {pages} {469} (\bibinfo {year} {1984})}\BibitemShut {NoStop}%
\bibitem [{\citenamefont {Olmsted}\ and\ \citenamefont
  {Goldbart}(1990)}]{Olmsted1990}%
  \BibitemOpen
  \bibfield  {author} {\bibinfo {author} {\bibfnamefont {P.~D.}\ \bibnamefont
  {Olmsted}}\ and\ \bibinfo {author} {\bibfnamefont {P.~M.}\ \bibnamefont
  {Goldbart}},\ }\href {\doibase 10.1103/PhysRevA.41.4578} {\bibfield
  {journal} {\bibinfo  {journal} {Physical Review A}\ }\textbf {\bibinfo
  {volume} {41}},\ \bibinfo {pages} {4578} (\bibinfo {year}
  {1990})}\BibitemShut {NoStop}%
\bibitem [{\citenamefont {Chen}\ \emph {et~al.}(2020)\citenamefont {Chen},
  \citenamefont {Chen}, \citenamefont {Korblova}, \citenamefont {Korblova},
  \citenamefont {Dong}, \citenamefont {Dong}, \citenamefont {Wei},
  \citenamefont {Wei}, \citenamefont {Shao}, \citenamefont {Shao},
  \citenamefont {Radzihovsky}, \citenamefont {Radzihovsky}, \citenamefont
  {Glaser}, \citenamefont {Glaser}, \citenamefont {MacLennan}, \citenamefont
  {MacLennan}, \citenamefont {Bedrov}, \citenamefont {Bedrov}, \citenamefont
  {Walba}, \citenamefont {Walba}, \citenamefont {Clark},\ and\ \citenamefont
  {Clark}}]{Chen2020}%
  \BibitemOpen
  \bibfield  {author} {\bibinfo {author} {\bibfnamefont {X.}~\bibnamefont
  {Chen}}, \bibinfo {author} {\bibfnamefont {X.}~\bibnamefont {Chen}}, \bibinfo
  {author} {\bibfnamefont {E.}~\bibnamefont {Korblova}}, \bibinfo {author}
  {\bibfnamefont {E.}~\bibnamefont {Korblova}}, \bibinfo {author}
  {\bibfnamefont {D.}~\bibnamefont {Dong}}, \bibinfo {author} {\bibfnamefont
  {D.}~\bibnamefont {Dong}}, \bibinfo {author} {\bibfnamefont {X.}~\bibnamefont
  {Wei}}, \bibinfo {author} {\bibfnamefont {X.}~\bibnamefont {Wei}}, \bibinfo
  {author} {\bibfnamefont {R.}~\bibnamefont {Shao}}, \bibinfo {author}
  {\bibfnamefont {R.}~\bibnamefont {Shao}}, \bibinfo {author} {\bibfnamefont
  {L.}~\bibnamefont {Radzihovsky}}, \bibinfo {author} {\bibfnamefont
  {L.}~\bibnamefont {Radzihovsky}}, \bibinfo {author} {\bibfnamefont {M.~A.}\
  \bibnamefont {Glaser}}, \bibinfo {author} {\bibfnamefont {M.~A.}\
  \bibnamefont {Glaser}}, \bibinfo {author} {\bibfnamefont {J.~E.}\
  \bibnamefont {MacLennan}}, \bibinfo {author} {\bibfnamefont {J.~E.}\
  \bibnamefont {MacLennan}}, \bibinfo {author} {\bibfnamefont {D.}~\bibnamefont
  {Bedrov}}, \bibinfo {author} {\bibfnamefont {D.}~\bibnamefont {Bedrov}},
  \bibinfo {author} {\bibfnamefont {D.~M.}\ \bibnamefont {Walba}}, \bibinfo
  {author} {\bibfnamefont {D.~M.}\ \bibnamefont {Walba}}, \bibinfo {author}
  {\bibfnamefont {N.~A.}\ \bibnamefont {Clark}}, \ and\ \bibinfo {author}
  {\bibfnamefont {N.~A.}\ \bibnamefont {Clark}},\ }\href {\doibase
  10.1073/pnas.2002290117} {\bibfield  {journal} {\bibinfo  {journal}
  {Proceedings of the National Academy of Sciences of the United States of
  America}\ }\textbf {\bibinfo {volume} {117}},\ \bibinfo {pages} {14021}
  (\bibinfo {year} {2020})},\ \Eprint {http://arxiv.org/abs/2003.03020}
  {arXiv:2003.03020} \BibitemShut {NoStop}%
\bibitem [{\citenamefont {Onsager}(1931)}]{Onsager1931b}%
  \BibitemOpen
  \bibfield  {author} {\bibinfo {author} {\bibfnamefont {L.}~\bibnamefont
  {Onsager}},\ }\href {\doibase 10.1103/PhysRev.38.2265} {\bibfield  {journal}
  {\bibinfo  {journal} {Physical Review}\ }\textbf {\bibinfo {volume} {38}},\
  \bibinfo {pages} {2265} (\bibinfo {year} {1931})}\BibitemShut {NoStop}%
\bibitem [{\citenamefont {Fang}\ \emph {et~al.}(2019)\citenamefont {Fang},
  \citenamefont {Kruse}, \citenamefont {Lu},\ and\ \citenamefont
  {Wang}}]{Fang2019}%
  \BibitemOpen
  \bibfield  {author} {\bibinfo {author} {\bibfnamefont {X.}~\bibnamefont
  {Fang}}, \bibinfo {author} {\bibfnamefont {K.}~\bibnamefont {Kruse}},
  \bibinfo {author} {\bibfnamefont {T.}~\bibnamefont {Lu}}, \ and\ \bibinfo
  {author} {\bibfnamefont {J.}~\bibnamefont {Wang}},\ }\href {\doibase
  10.1103/RevModPhys.91.045004} {\bibfield  {journal} {\bibinfo  {journal}
  {Reviews of Modern Physics}\ }\textbf {\bibinfo {volume} {91}},\ \bibinfo
  {pages} {45004} (\bibinfo {year} {2019})},\ \Eprint
  {http://arxiv.org/abs/2012.05067} {arXiv:2012.05067} \BibitemShut {NoStop}%
\bibitem [{\citenamefont {Plischke}\ and\ \citenamefont
  {Bergersen}(2005)}]{Plischke2005}%
  \BibitemOpen
  \bibfield  {author} {\bibinfo {author} {\bibfnamefont {M.}~\bibnamefont
  {Plischke}}\ and\ \bibinfo {author} {\bibfnamefont {B.}~\bibnamefont
  {Bergersen}},\ }\href@noop {} {\emph {\bibinfo {title} {Physics}}}\ (\bibinfo
  {year} {2005})\BibitemShut {NoStop}%
\bibitem [{\citenamefont {Bialek}\ \emph {et~al.}(2012)\citenamefont {Bialek},
  \citenamefont {Cavagna}, \citenamefont {Giardina}, \citenamefont {Mora},
  \citenamefont {Silvestri}, \citenamefont {Viale},\ and\ \citenamefont
  {Walczak}}]{Bialek2011}%
  \BibitemOpen
  \bibfield  {author} {\bibinfo {author} {\bibfnamefont {W.}~\bibnamefont
  {Bialek}}, \bibinfo {author} {\bibfnamefont {A.}~\bibnamefont {Cavagna}},
  \bibinfo {author} {\bibfnamefont {I.}~\bibnamefont {Giardina}}, \bibinfo
  {author} {\bibfnamefont {T.}~\bibnamefont {Mora}}, \bibinfo {author}
  {\bibfnamefont {E.}~\bibnamefont {Silvestri}}, \bibinfo {author}
  {\bibfnamefont {M.}~\bibnamefont {Viale}}, \ and\ \bibinfo {author}
  {\bibfnamefont {A.~M.}\ \bibnamefont {Walczak}},\ }\href {\doibase
  10.1073/pnas.1118633109} {\bibfield  {journal} {\bibinfo  {journal}
  {Proceedings of the National Academy of Sciences}\ }\textbf {\bibinfo
  {volume} {109}},\ \bibinfo {pages} {4786} (\bibinfo {year} {2012})},\ \Eprint
  {http://arxiv.org/abs/1107.0604} {arXiv:1107.0604} \BibitemShut {NoStop}%
\bibitem [{\citenamefont {Athilingam}\ \emph {et~al.}(2021)\citenamefont
  {Athilingam}, \citenamefont {Tiwari}, \citenamefont {Toyama},\ and\
  \citenamefont {Saunders}}]{athilingam2021mechanics}%
  \BibitemOpen
  \bibfield  {author} {\bibinfo {author} {\bibfnamefont {T.}~\bibnamefont
  {Athilingam}}, \bibinfo {author} {\bibfnamefont {P.}~\bibnamefont {Tiwari}},
  \bibinfo {author} {\bibfnamefont {Y.}~\bibnamefont {Toyama}}, \ and\ \bibinfo
  {author} {\bibfnamefont {T.~E.}\ \bibnamefont {Saunders}},\ }in\ \href@noop
  {} {\emph {\bibinfo {booktitle} {Seminars in Cell \& Developmental
  Biology}}},\ Vol.\ \bibinfo {volume} {120}\ (\bibinfo {organization}
  {Elsevier},\ \bibinfo {year} {2021})\ pp.\ \bibinfo {pages}
  {171--180}\BibitemShut {NoStop}%
\bibitem [{\citenamefont {Aigouy}\ \emph
  {et~al.}(2010{\natexlab{b}})\citenamefont {Aigouy}, \citenamefont
  {Farhadifar}, \citenamefont {Staple}, \citenamefont {Sagner}, \citenamefont
  {R{\"o}per}, \citenamefont {J{\"u}licher},\ and\ \citenamefont
  {Eaton}}]{aigouy2010cell}%
  \BibitemOpen
  \bibfield  {author} {\bibinfo {author} {\bibfnamefont {B.}~\bibnamefont
  {Aigouy}}, \bibinfo {author} {\bibfnamefont {R.}~\bibnamefont {Farhadifar}},
  \bibinfo {author} {\bibfnamefont {D.~B.}\ \bibnamefont {Staple}}, \bibinfo
  {author} {\bibfnamefont {A.}~\bibnamefont {Sagner}}, \bibinfo {author}
  {\bibfnamefont {J.-C.}\ \bibnamefont {R{\"o}per}}, \bibinfo {author}
  {\bibfnamefont {F.}~\bibnamefont {J{\"u}licher}}, \ and\ \bibinfo {author}
  {\bibfnamefont {S.}~\bibnamefont {Eaton}},\ }\href@noop {} {\bibfield
  {journal} {\bibinfo  {journal} {Cell}\ }\textbf {\bibinfo {volume} {142}},\
  \bibinfo {pages} {773} (\bibinfo {year} {2010}{\natexlab{b}})}\BibitemShut
  {NoStop}%
\bibitem [{\citenamefont {Liverpool}(2020)}]{Liverpool2020}%
  \BibitemOpen
  \bibfield  {author} {\bibinfo {author} {\bibfnamefont {T.~B.}\ \bibnamefont
  {Liverpool}},\ }\href {\doibase 10.1103/PhysRevE.101.042107} {\bibfield
  {journal} {\bibinfo  {journal} {Physical Review E}\ }\textbf {\bibinfo
  {volume} {101}},\ \bibinfo {pages} {042107} (\bibinfo {year}
  {2020})}\BibitemShut {NoStop}%
\bibitem [{\citenamefont {Raser}\ and\ \citenamefont
  {O'Shea}(2005)}]{Raser2005}%
  \BibitemOpen
  \bibfield  {author} {\bibinfo {author} {\bibfnamefont {J.~M.}\ \bibnamefont
  {Raser}}\ and\ \bibinfo {author} {\bibfnamefont {E.~K.}\ \bibnamefont
  {O'Shea}},\ }\href {\doibase 10.1126/science.1105891} {\bibfield  {journal}
  {\bibinfo  {journal} {Science (New York, N.Y.)}\ }\textbf {\bibinfo {volume}
  {309}},\ \bibinfo {pages} {2010} (\bibinfo {year} {2005})}\BibitemShut
  {NoStop}%
\bibitem [{\citenamefont {Razzell}\ \emph {et~al.}(2011)\citenamefont
  {Razzell}, \citenamefont {Wood},\ and\ \citenamefont
  {Martin}}]{razzell2011swatting}%
  \BibitemOpen
  \bibfield  {author} {\bibinfo {author} {\bibfnamefont {W.}~\bibnamefont
  {Razzell}}, \bibinfo {author} {\bibfnamefont {W.}~\bibnamefont {Wood}}, \
  and\ \bibinfo {author} {\bibfnamefont {P.}~\bibnamefont {Martin}},\
  }\href@noop {} {\bibfield  {journal} {\bibinfo  {journal} {Disease models \&
  mechanisms}\ }\textbf {\bibinfo {volume} {4}},\ \bibinfo {pages} {569}
  (\bibinfo {year} {2011})}\BibitemShut {NoStop}%
\bibitem [{\citenamefont {Rueden}\ \emph {et~al.}(2017)\citenamefont {Rueden},
  \citenamefont {Schindelin}, \citenamefont {Hiner}, \citenamefont {DeZonia},
  \citenamefont {Walter}, \citenamefont {Arena},\ and\ \citenamefont
  {Eliceiri}}]{rueden2017imagej2}%
  \BibitemOpen
  \bibfield  {author} {\bibinfo {author} {\bibfnamefont {C.~T.}\ \bibnamefont
  {Rueden}}, \bibinfo {author} {\bibfnamefont {J.}~\bibnamefont {Schindelin}},
  \bibinfo {author} {\bibfnamefont {M.~C.}\ \bibnamefont {Hiner}}, \bibinfo
  {author} {\bibfnamefont {B.~E.}\ \bibnamefont {DeZonia}}, \bibinfo {author}
  {\bibfnamefont {A.~E.}\ \bibnamefont {Walter}}, \bibinfo {author}
  {\bibfnamefont {E.~T.}\ \bibnamefont {Arena}}, \ and\ \bibinfo {author}
  {\bibfnamefont {K.~W.}\ \bibnamefont {Eliceiri}},\ }\href@noop {} {\bibfield
  {journal} {\bibinfo  {journal} {BMC bioinformatics}\ }\textbf {\bibinfo
  {volume} {18}},\ \bibinfo {pages} {529} (\bibinfo {year} {2017})}\BibitemShut
  {NoStop}%
\bibitem [{\citenamefont {Schindelin}\ \emph {et~al.}(2012)\citenamefont
  {Schindelin}, \citenamefont {Arganda-Carreras}, \citenamefont {Frise},
  \citenamefont {Kaynig}, \citenamefont {Longair}, \citenamefont {Pietzsch},
  \citenamefont {Preibisch}, \citenamefont {Rueden}, \citenamefont {Saalfeld},
  \citenamefont {Schmid} \emph {et~al.}}]{schindelin2012fiji}%
  \BibitemOpen
  \bibfield  {author} {\bibinfo {author} {\bibfnamefont {J.}~\bibnamefont
  {Schindelin}}, \bibinfo {author} {\bibfnamefont {I.}~\bibnamefont
  {Arganda-Carreras}}, \bibinfo {author} {\bibfnamefont {E.}~\bibnamefont
  {Frise}}, \bibinfo {author} {\bibfnamefont {V.}~\bibnamefont {Kaynig}},
  \bibinfo {author} {\bibfnamefont {M.}~\bibnamefont {Longair}}, \bibinfo
  {author} {\bibfnamefont {T.}~\bibnamefont {Pietzsch}}, \bibinfo {author}
  {\bibfnamefont {S.}~\bibnamefont {Preibisch}}, \bibinfo {author}
  {\bibfnamefont {C.}~\bibnamefont {Rueden}}, \bibinfo {author} {\bibfnamefont
  {S.}~\bibnamefont {Saalfeld}}, \bibinfo {author} {\bibfnamefont
  {B.}~\bibnamefont {Schmid}},  \emph {et~al.},\ }\href@noop {} {\bibfield
  {journal} {\bibinfo  {journal} {Nature methods}\ }\textbf {\bibinfo {volume}
  {9}},\ \bibinfo {pages} {676} (\bibinfo {year} {2012})}\BibitemShut {NoStop}%
\bibitem [{\citenamefont {Cross}(2018)}]{stephen_cross_2018_2525263}%
  \BibitemOpen
  \bibfield  {author} {\bibinfo {author} {\bibfnamefont {S.}~\bibnamefont
  {Cross}},\ }\href@noop {} {\enquote {\bibinfo {title}
  {Sjcross/modularimageanalysis: Version 0.7.21},}\ } (\bibinfo {year}
  {2018}),\ \bibinfo {note} {dOI: 10.5281/zenodo.2525263}\BibitemShut {NoStop}%
\bibitem [{\citenamefont {Umorin}(2002)}]{Umorin_2002}%
  \BibitemOpen
  \bibfield  {author} {\bibinfo {author} {\bibfnamefont {M.}~\bibnamefont
  {Umorin}},\ }\href@noop {} {\enquote {\bibinfo {title} {Stack focuser plugin
  for imagej},}\ }\bibinfo {howpublished}
  {https://imagej.nih.gov/ij/plugins/stack-focuser.html} (\bibinfo {year}
  {2002})\BibitemShut {NoStop}%
\bibitem [{\citenamefont {Thuma}\ \emph {et~al.}(2018)\citenamefont {Thuma},
  \citenamefont {Carter}, \citenamefont {Weavers},\ and\ \citenamefont
  {Martin}}]{thuma2018drosophila}%
  \BibitemOpen
  \bibfield  {author} {\bibinfo {author} {\bibfnamefont {L.}~\bibnamefont
  {Thuma}}, \bibinfo {author} {\bibfnamefont {D.}~\bibnamefont {Carter}},
  \bibinfo {author} {\bibfnamefont {H.}~\bibnamefont {Weavers}}, \ and\
  \bibinfo {author} {\bibfnamefont {P.}~\bibnamefont {Martin}},\ }\href@noop {}
  {\bibfield  {journal} {\bibinfo  {journal} {Journal of Cell Biology}\
  }\textbf {\bibinfo {volume} {217}},\ \bibinfo {pages} {3045} (\bibinfo {year}
  {2018})}\BibitemShut {NoStop}%
\bibitem [{\citenamefont {Arganda-Carreras}\ \emph {et~al.}(2016)\citenamefont
  {Arganda-Carreras}, \citenamefont {Kaynig}, \citenamefont {Rueden},
  \citenamefont {Schindelin}, \citenamefont {Cardona},\ and\ \citenamefont
  {Seung}}]{arganda_carreras_2016_59290}%
  \BibitemOpen
  \bibfield  {author} {\bibinfo {author} {\bibfnamefont {I.}~\bibnamefont
  {Arganda-Carreras}}, \bibinfo {author} {\bibfnamefont {V.}~\bibnamefont
  {Kaynig}}, \bibinfo {author} {\bibfnamefont {C.}~\bibnamefont {Rueden}},
  \bibinfo {author} {\bibfnamefont {J.}~\bibnamefont {Schindelin}}, \bibinfo
  {author} {\bibfnamefont {A.}~\bibnamefont {Cardona}}, \ and\ \bibinfo
  {author} {\bibfnamefont {H.~S.}\ \bibnamefont {Seung}},\ }\href@noop {}
  {\enquote {\bibinfo {title} {Trainable\_segmentation: Release v3.1.2},}\ }
  (\bibinfo {year} {2016}),\ \bibinfo {note} {dOI:
  10.5281/zenodo.59290}\BibitemShut {NoStop}%
\bibitem [{\citenamefont {Steger}(1996)}]{steger1996calculation}%
  \BibitemOpen
  \bibfield  {author} {\bibinfo {author} {\bibfnamefont {C.}~\bibnamefont
  {Steger}},\ }\href@noop {} {\bibfield  {journal} {\bibinfo  {journal}
  {Munchen Univ., Munchen, Germany, Tech. Rep. FGBV-96-05}\ } (\bibinfo {year}
  {1996})}\BibitemShut {NoStop}%
\bibitem [{\citenamefont {Graner}\ \emph {et~al.}(2008)\citenamefont {Graner},
  \citenamefont {Dollet}, \citenamefont {Raufaste},\ and\ \citenamefont
  {Marmottant}}]{Graner2008}%
  \BibitemOpen
  \bibfield  {author} {\bibinfo {author} {\bibfnamefont {F.}~\bibnamefont
  {Graner}}, \bibinfo {author} {\bibfnamefont {B.}~\bibnamefont {Dollet}},
  \bibinfo {author} {\bibfnamefont {C.}~\bibnamefont {Raufaste}}, \ and\
  \bibinfo {author} {\bibfnamefont {P.}~\bibnamefont {Marmottant}},\ }\href
  {\doibase 10.1140/epje/i2007-10298-8} {\bibfield  {journal} {\bibinfo
  {journal} {The European Physical Journal E}\ }\textbf {\bibinfo {volume}
  {25}},\ \bibinfo {pages} {349} (\bibinfo {year} {2008})}\BibitemShut
  {NoStop}%
\bibitem [{\citenamefont {Marmottant}\ \emph {et~al.}(2008)\citenamefont
  {Marmottant}, \citenamefont {Raufaste},\ and\ \citenamefont
  {Graner}}]{Marmottant2008}%
  \BibitemOpen
  \bibfield  {author} {\bibinfo {author} {\bibfnamefont {P.}~\bibnamefont
  {Marmottant}}, \bibinfo {author} {\bibfnamefont {C.}~\bibnamefont
  {Raufaste}}, \ and\ \bibinfo {author} {\bibfnamefont {F.}~\bibnamefont
  {Graner}},\ }\href {\doibase 10.1140/epje/i2007-10300-7} {\bibfield
  {journal} {\bibinfo  {journal} {The European Physical Journal E}\ }\textbf
  {\bibinfo {volume} {25}},\ \bibinfo {pages} {371} (\bibinfo {year} {2008})},\
  \Eprint {http://arxiv.org/abs/0609188} {arXiv:0609188 [cond-mat]}
  \BibitemShut {NoStop}%
\bibitem [{\citenamefont {Esposito}(2012)}]{Esposito2012}%
  \BibitemOpen
  \bibfield  {author} {\bibinfo {author} {\bibfnamefont {M.}~\bibnamefont
  {Esposito}},\ }\href@noop {} {\bibfield  {journal} {\bibinfo  {journal}
  {Physical Review E}\ }\textbf {\bibinfo {volume} {85}},\ \bibinfo {pages}
  {041125} (\bibinfo {year} {2012})}\BibitemShut {NoStop}%
\bibitem [{\citenamefont {Zwanzig}(1960)}]{Zwanzig1960}%
  \BibitemOpen
  \bibfield  {author} {\bibinfo {author} {\bibfnamefont {R.}~\bibnamefont
  {Zwanzig}},\ }\href {\doibase 10.1063/1.1731409} {\bibfield  {journal}
  {\bibinfo  {journal} {The Journal of Chemical Physics}\ }\textbf {\bibinfo
  {volume} {33}},\ \bibinfo {pages} {1338} (\bibinfo {year}
  {1960})}\BibitemShut {NoStop}%
\bibitem [{\citenamefont {Voliotis}\ \emph {et~al.}(2009)\citenamefont
  {Voliotis}, \citenamefont {Cohen}, \citenamefont {Molina-Par{\'\i}s},\ and\
  \citenamefont {Liverpool}}]{Voliotis2009}%
  \BibitemOpen
  \bibfield  {author} {\bibinfo {author} {\bibfnamefont {M.}~\bibnamefont
  {Voliotis}}, \bibinfo {author} {\bibfnamefont {N.}~\bibnamefont {Cohen}},
  \bibinfo {author} {\bibfnamefont {C.}~\bibnamefont {Molina-Par{\'\i}s}}, \
  and\ \bibinfo {author} {\bibfnamefont {T.~B.}\ \bibnamefont {Liverpool}},\
  }\href@noop {} {\bibfield  {journal} {\bibinfo  {journal} {Physical review
  letters}\ }\textbf {\bibinfo {volume} {102}},\ \bibinfo {pages} {258101}
  (\bibinfo {year} {2009})}\BibitemShut {NoStop}%
\bibitem [{\citenamefont {Mogilner}\ and\ \citenamefont
  {Craig}(2010)}]{Mogilner2010}%
  \BibitemOpen
  \bibfield  {author} {\bibinfo {author} {\bibfnamefont {A.}~\bibnamefont
  {Mogilner}}\ and\ \bibinfo {author} {\bibfnamefont {E.}~\bibnamefont
  {Craig}},\ }\href@noop {} {\bibfield  {journal} {\bibinfo  {journal} {Journal
  of cell science}\ }\textbf {\bibinfo {volume} {123}},\ \bibinfo {pages}
  {3435} (\bibinfo {year} {2010})}\BibitemShut {NoStop}%
\bibitem [{\citenamefont {Sartori}\ and\ \citenamefont
  {Pigolotti}(2013)}]{Sartori2013}%
  \BibitemOpen
  \bibfield  {author} {\bibinfo {author} {\bibfnamefont {P.}~\bibnamefont
  {Sartori}}\ and\ \bibinfo {author} {\bibfnamefont {S.}~\bibnamefont
  {Pigolotti}},\ }\href@noop {} {\bibfield  {journal} {\bibinfo  {journal}
  {Physical review letters}\ }\textbf {\bibinfo {volume} {110}},\ \bibinfo
  {pages} {188101} (\bibinfo {year} {2013})}\BibitemShut {NoStop}%
\end{thebibliography}%



\end{document}